\documentclass[a4paper,11pt]{article}
\usepackage{amsmath}
\usepackage{amssymb}
\usepackage{graphicx}
\usepackage{color}
\usepackage[square,numbers,comma,sort&compress]{natbib}
\usepackage[colorlinks=true,citecolor=blue,linkcolor=blue,urlcolor=blue]{hyperref}
\usepackage{geometry}
\usepackage[normalem]{ulem}
\def\varabstract{ }
\def\varkeywords{ }
\def\vararxivnumber{ }
\def\vartitle{ }
\def\varpreprint{ }
\renewcommand{\title}[1]{\gdef\vartitle{#1}}
\renewcommand{\abstract}[1]{\gdef\varabstract{#1}}
\newcommand{\keywords}[1]{\gdef\varkeywords{#1}}
\newcommand{\arxivnumber}[1]{\gdef\vararxivnumber{#1}}
\newcommand{\preprint}[1]{\gdef\varpreprint{#1}}
\newtoks\authtoks
\renewcommand{\author}[2][]{%
	\authtoks=\expandafter{\the\authtoks#2$^{#1}$\ }%
}
\newtoks\affiltoks
\newcommand{\affiliation}[2][]{%
    \affiltoks=\expandafter{\the\affiltoks{\item[$^{#1}$]#2}}%
}
\newtoks\emailtoks\newcounter{emailcounter}%
\newcommand{\emailAdd}[1]{%
\ifnum\theemailcounter>0\emailtoks=\expandafter{\the\emailtoks, \typeemail{#1}}%
\else\emailtoks=\expandafter{\typeemail{#1}}%
\fi
\stepcounter{emailcounter}}
\newcommand{\typeemail}[1]{\href{mailto:#1}{\tt #1}}

\renewcommand\maketitle{
	\newgeometry{margin=2cm}
	\pagestyle{empty}\setcounter{page}{0}
	\if!\varpreprint!\else\begin{flushright}\varpreprint\end{flushright}\fi
	{\huge\flushleft\sffamily\bfseries\vartitle\par}
\vskip6ex
{\large\bfseries\raggedright\sffamily\the\authtoks\par}
\vskip2ex
\begin{list}{}{%
\setlength{\leftmargin}{0.28cm}%
\setlength{\labelsep}{0pt}%
\setlength{\itemsep}{-3pt}%
\setlength{\topsep}{-\parskip}}
\itshape\small%
\the\affiltoks
\end{list}
\vskip2ex
\noindent\hspace{0.28cm}\begin{minipage}[l]{.9\textwidth}
\begin{flushleft}
\textit{E-mail:} \the\emailtoks
\end{flushleft}
\end{minipage}
\vskip5ex
\noindent{\renewcommand\baselinestretch{.9}\textsc{Abstract:}}\ \varabstract
\vskip5ex 
\if!\varkeywords!\else\noindent{\textsc{Keywords:}}\ \varkeywords \vskip2ex\fi
\if!\vararxivnumber!\else\noindent{\textsc{ArXiv ePrint:}} \href{http://arxiv.org/abs/\vararxivnumber}{\vararxivnumber}\vskip2ex\fi
\newpage
\restoregeometry
\pagestyle{plain}
\hrule
\bigskip\bigskip

{
	\hypersetup{linkcolor=black}
	\tableofcontents
}
\bigskip\medskip
\hrule
\bigskip\bigskip
\setcounter{footnote}{0}
} 

\usepackage[utf8]{inputenc}
\usepackage{verbatim}
\usepackage{textcomp}
\usepackage{nicefrac}
\usepackage{subcaption}
\usepackage{diagbox}
\usepackage{booktabs}

\usepackage{array}
\usepackage{multirow}
\usepackage{slashed}

\definecolor{MS}{rgb}{1,0,0}                                                                   
\definecolor{all}{rgb}{1,0,1}

\newcommand{\braket}[1]{\ensuremath{\left\langle#1\right\rangle}}

\def\be{\begin{equation}}
\def\ee{\end{equation}}

\title{Dark matter direct detection of a fermionic singlet at one loop}

\author[a]{Juan Herrero-Garc\'ia,}
\author[b,c]{Emiliano Molinaro}
\author[d]{and Michael A.~Schmidt}
\affiliation[a]{ARC Center of Excellence for Particle Physics at the Terascale, University of Adelaide, 5005 Adelaide, South Australia, Australia}
\affiliation[b]{Department of Physics and Astronomy, University of Aarhus, Ny Munkegade 120, DK-8000 Aarhus C, Denmark}
\affiliation[c]{CP$^3$-Origins, University of Southern Denmark, Campusvej 55, 5230 Odense M, Denmark}
\affiliation[d]{ARC Centre of Excellence for Particle Physics at the Terascale, School of Physics, The University of Sydney, Physics Road, 2006, New South Wales, Australia}

\emailAdd{juan.herrero-garcia@adelaide.edu.au}
\emailAdd{molinaro@phys.au.dk}
\emailAdd{michael.schmidt@sydney.edu.au}

\abstract{The strong direct detection limits could be pointing to dark matter -- nucleus scattering at loop level. We study in detail the prototype
example of an electroweak singlet (Dirac or Majorana) dark matter fermion coupled to an extended
dark sector, which is composed of a new fermion and a new scalar. Given the strong
limits on colored particles from direct and indirect searches we assume that
the fields of the new dark sector are color singlets. We outline the possible simplified models, including the well-motivated cases in which the extra
scalar or fermion is a Standard Model particle, as well as the possible connection to neutrino masses.
We compute the
contributions to direct detection from the photon, the $Z$ and the Higgs penguins for arbitrary
quantum numbers of the dark sector.  Furthermore, we derive
compact expressions in certain limits, i.e., when all new particles are heavier
than the dark matter mass and when the fermion running in the loop is light, like a Standard Model lepton. We study in detail the predicted direct detection rate and how
current and future direct detection limits constrain the model parameters. In case dark matter couples directly to Standard Model leptons we find an interesting interplay between lepton flavor violation, direct detection and the observed relic abundance.
}

\arxivnumber{1803.05660}
\preprint{ADP-18-8/T1056\\
CP3-Origins-2018-009-DNRF90}
\makeatother

\begin{document}

\maketitle
\section{Introduction} \label{sec:intro}
Direct detection (DD) experiments search for dark matter (DM) scatterings off nuclei in underground detectors. The current limits impose very strong constraints on the
parameters of Weakly Interacting Massive Particles (WIMPs), which
are one of the prototype DM candidates. The current most
stringent DD limits for WIMPs in the mass range of $[10,1000]$
GeV come from xenon experiments~\cite{Akerib:2016vxi,Cui:2017nnn,Aprile:2017iyp}. In
this work we hypothesize that the absence of DD signals may be
reconciled with the WIMP paradigm by generating the scattering at
one-loop order and thus with an extra $1/(16\pi^2)^2$ suppression
of the cross section. As we will see, current and next-generation
experiments are able to test significant regions in parameter
space of this class of scenarios.

There have been several works in the literature on DD at one-loop order. In
Refs.~\cite{Kopp:2009et, Agrawal:2011ze, Kopp:2014tsa,Baek:2015fma} the authors studied DD
limits from photon interactions in the context of flavored DM and in
Ref.~\cite{Schmidt:2012yg} in the context of a radiative neutrino mass model
(the scotogenic model~\cite{Ma:2006km}) with inelastic Majorana DM. In
Ref.~\cite{Ibarra:2015fqa} the authors performed a detailed study of one-loop
scenarios with a charged mediator directly coupled to Standard Model (SM)
fields, including the $Z$ and Higgs boson contributions. For couplings to the first and second generation of quarks the dominant contribution may be due to scattering at tree level, while box diagrams may be significant for third generation quarks. Similarly, Ref.~\cite{Sandick:2016zut} studied direct detection of Majorana DM directly coupled to both left- and right-handed SM leptons via two charged scalar mediators. The $Z$ and Higgs contributions were also computed for the scotogenic model in Ref.~\cite{Ibarra:2016dlb} and also for
DM connected to the SM via a neutrino-portal in Ref.~\cite{Gonzalez-Macias:2016vxy}. In Ref.~\cite{Sanderson:2018lmj} the
authors studied the one-loop contributions to DD in models with pseudo-scalar
mediators or inelastic scattering. In the context of supersymmetry detailed
computations have been performed for the bino~\cite{Berlin:2015njh} and wino~\cite{Hisano:2004pv,Hisano:2010fy,Hisano:2015rsa} DM cases. In the latter scenario loop contributions to DM-nucleus scattering
due to gauge bosons may give significant corrections.

\begin{figure}\centering
	\includegraphics[width=0.4\linewidth]{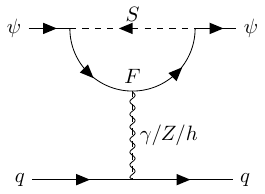}\hspace{1cm}\includegraphics[width=0.4\linewidth]{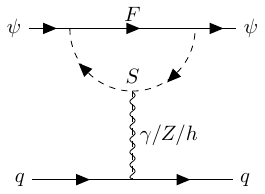}
\caption{One-loop penguin diagrams for fermionic singlet DM scattering off nuclei. They are generated with up to two heavy particles from a dark sector (a scalar $S$ and a fermion $F$). The photon and $Z$ boson are coupled to the new fermion (left diagram) or the new scalar (right diagram). For minimal models with one fermion the Higgs boson $h$ only couples to the scalar $S$, but SM fermions in the loop also lead to a Higgs penguin diagram where the SM Higgs boson is attached to the fermion line.
The possible quantum numbers of the dark particles are given in Tab.~\ref{tab:dark_sector}.}\label{fig:penguins}
\end{figure}

In this work we study the DD scattering rate for the case of DM being a SM singlet Dirac or Majorana fermion $\psi$, which is coupled to a
more complex dark sector. A conserved global U(1) or $Z_2$ symmetry is assumed in order to stabilize the DM particle. In our scenario there are no tree-level
contributions to the DD cross section. The lowest order
scattering off nuclei occurs at one-loop order via the penguin diagrams in Fig.~\ref{fig:penguins}, with a dark fermion $F$ and a dark scalar $S$ running in the loop.
We assume that the new particles are color singlets, so that there are no
flavor changing neutral currents in the quark sector, and there are only weak
limits from direct production at the Large Hadron Collider (LHC). In this way
box-diagram contributions to the scattering amplitude are absent. Our main goal
is to  study analytically the different contributions to the DM-nucleus scattering, as well as to outline possible  simplified models, including those with SM fields. In addition we analyze the current limits from DD, as
well as constraints coming from the relic abundance, lepton flavor violation (LFV) and anomalous magnetic dipole moments (AMMs).\footnote{In our scenario leptonic electric dipole moments appear only at two-loop order and are therefore suppressed.}

The paper is structured as follows: In Sec.~\ref{sec:setup} we study the UV
completions of the fermionic DM scenario including models with SM particles in
the loop. In order to fix the notation we review in Sec.~\ref{sec:opseft} the
relevant effective operators for DD at the quark level and also their
non-relativistic (NR) versions at the nucleon level. In Sec.~\ref{sec:computation}
we derive analytical expressions for the Wilson coefficients and provide
compact expressions in certain limits. In Sec.~\ref{sec:numerics} we perform a
numerical analysis of the phenomenology relevant for DD. First we show some
numerical examples for the Wilson coefficients at the quark and nucleon level
(the latter in their NR version). Afterwards we derive the
current limits on the model parameters and discuss future expected sensitivity.
We also discuss limits from LFV processes for models in which DM is
directly coupled to SM leptons. Sections~\ref{sec:computation} and~\ref{sec:numerics} contain the main results of this paper. We discuss other
phenomenological aspects of the proposed scenario, such as the DM relic
abundance, invisible decays and 
searches at colliders in Sec.~\ref{sec:other}. Finally we present our
conclusions in Sec.~\ref{sec:conc}. 

The manuscript also includes several appendices with technical details. The generalization to larger symmetry groups in the dark sector is presented in App.~\ref{ap:darkmattergroup}. In App.~\ref{ap:DD} we show a compact expression for the differential cross section in order to make contact with the literature and we briefly review the
differential event rate for DD. In App.~\ref{ap:HZdecays} we give generic expressions for Higgs and $Z$ boson invisible decays into DM. Relevant formulae for LFV observables and for AMMs of leptons are provided in App.~\ref{ap:LFV}. Details about the calculation of the relic abundance are collected in App.~\ref{ap:relic_details} and the numerical expressions for the matching to NR operators are given in App.~\ref{ap:NRmatching}.

\begin{table} [t]
\centering
\begin{tabular}{l|c|ccc|c}
\toprule
Dark sector &Field & SU(3)$_{\rm C}$ & SU(2)$_{\rm L}$ & U(1)$_{\rm Y}$ & U(1)$_{\rm dm}$\\
\midrule
Dark matter&$\psi$ & 1 & 1 & 0 & $ 1$\\ 
\midrule
Dark scalar&$S$ & 1 & $d_F$ & $Y_F$ & $q_s $\\ 
Dark fermion &$F$ & 1 & $d_F$ & $Y_F$ & $ q_s+1$\\ 
\bottomrule
\end{tabular}
\caption{Particle content and quantum numbers of the fermionic DM scenario with a dark fermion and a dark scalar.  The dark sector is charged under a global U(1) symmetry which stabilizes the DM $\psi$. 
}\label{tab:dark_sector}
\end{table}

\section{Fermionic singlet dark matter} \label{sec:setup}

In the following sections we first present simplified models of Dirac and Majorana fermion DM with vector-like fermions in the loop and then discuss SM particles in the loop.

\subsection{Dirac dark matter} 
The new particles can have different combinations of quantum numbers as displayed in Tab.~\ref{tab:dark_sector}. We consider a global U(1)$_{\rm dm}$ symmetry in the dark sector to stabilize DM. It can equally be replaced by a discrete $Z_n$ subgroup. Other symmetry groups are discussed in App.~\ref{ap:darkmattergroup}.

The interaction Lagrangian for the fields $\psi$, $F$ and $S$ reads
\begin{eqnarray}
\begin{split}
\mathcal{L}_{\psi} = &\; i\, \overline \psi\, \slashed \partial \,\psi\,-\, m_{\psi}\, \overline{\psi}\, \psi \,+\,  i\, \overline F\, \slashed D \,F \, -\, m_{F}\, \overline{F}\, F + \left( D_\mu S\right)^\dagger D^\mu S - \mathcal{V}(S,H) \\
    &  -\, \Big( y_1 \, \overline{F_{\rm R}} \, S\,\psi_{\rm L}\,+y_2 \, \overline{F_{\rm L}} \, S\,\psi_{\rm R}\,+\,\text{H.c.}\Big)\,, \label{Lpsid}
\end{split}
\end{eqnarray}
where $H$ is the SM Higgs doublet\footnote{We define the SM Higgs doublet $H$ with hypercharge $1/2$.} and $\mathcal{V}(S,H)$ denotes
the scalar potential.
The DM $\psi$ is a SM fermion singlet, but is charged under the dark sector symmetry. The fields $F$ and $S$ are charged under
the electroweak gauge group. Electroweak gauge invariance
requires them to be in the same SU(2)$_\text{L}$ irreducible
representation of dimension $d_F$ and to have equal hypercharge
$Y_F$.
Notice that in some cases there can be interactions with the SM fields which are subject to strong constraints. We discuss such cases in Sec.~\eqref{sec:darktoSM}.

In the case of a global symmetry, even if DM is stable at the renormalizable level,
higher-order Planck-scale suppressed operators may induce its
decay~\cite{Mambrini:2015sia}. In particular for a Dirac fermion $\psi$ the
dimension-5 operator $\overline \psi {\tilde H}^\dagger (\slashed{D} L)$ with the SM
lepton doublet $L$ is one such example. One
can construct UV completions of such operators by softly-breaking the global
symmetry in the dark sector which induces decays, possibly radiatively.
The limits dramatically depend on the DM mass and the Wilson coefficient of the
operator. For the rest of the paper we assume that DM is cosmologically stable and that it satisfies all indirect detection constraints on decaying DM.

In our simplified scenario with  the interactions given in
Eq.~\eqref{Lpsid} it would seem that two of the three new states were stable:
$\psi$ and one of $S$ or $F$. For the following discussion let us assume $m_S \geq
m_F+m_\psi$ so that $F$ is potentially stable, while $S$ can decay.\footnote{Similar arguments apply to the other case where $S$ might be stable.} Then, there are two possibilities:
(i) If the fermion $F$ is a SM singlet (but charged under the dark group), it also contributes to the DM relic density.\footnote{In this case, if $m_\psi\simeq m_F$, coannihilations play an important role \cite{Griest:1990kh}.} Hence, the DD rate of the $\psi$ has to be rescaled by its
smaller density under the assumption that the global density scales as the local one, and there is a similar DD rate for $F$ via Higgs penguins with $\psi$ in the loop. 
(ii) If $F$ is charged under the SM group (SU(2)$_\text{L}$ charges and/or
	hypercharge) its electrically charged components have to decay
	given the stringent limits on charged stable particles~\cite{Davidson:2000hf,Kudo:2001ie,McDermott:2010pa,SanchezSalcedo:2010ev}. If the components of $F$ mix with SM leptons, they decay like in the
	model discussed in Refs.~\cite{Dissauer:2012xa, Frandsen:2013bfa}.
	Otherwise, as for the DM via the interaction $\overline \psi {\tilde H}^\dagger (\slashed{D} L)$, the fermion  $F$ may also decay into SM particles via
	non-renormalizable operators, which are allowed on general grounds, unless $F$ carries fractional electric charge, or other symmetries forbid them. In this case the fermion $F$ has to decay much faster than the long-lived DM particle.\footnote{Naively, the scalar $S$ being lighter than the fermion $F$ appears to be more natural given that there are 13 dimension-5 operators which induce decay for a scalar compared to one for a fermion~\cite{Mambrini:2015sia}.
} 

If $F$ is a SM lepton, a charged lepton or a neutrino, it may be stable.
Similarly, if $F$ is a right-handed neutrino, it mixes with SM neutrinos and decay.  Also, in the case in which $S$ is the SM Higgs doublet and $F$ a heavier fermion, the latter may decay into the SM Higgs boson. We discuss all these possibilities in more detail below. 

In general the SM Higgs boson couples to the new scalar multiplet $S$ via a Higgs portal interaction in the scalar potential $\mathcal{V}(S,H)$. Depending on the quantum numbers of the particles in the dark sector, it may also have an interaction with the fermion $F$, for instance if the latter is a SM lepton. In the case of a charged lepton $\ell$ in the loop, the largest Higgs interactions are proportional to the square of its mass $(m_\ell/v)^2\ll 1$, with the electroweak vacuum expectation value (VEV) $v\simeq 246$ GeV. Therefore this contribution is suppressed and can be safely neglected. While these interactions are even further suppressed for Dirac neutrinos, in principle it is possible to have $\mathcal{O}(1)$ Yukawa couplings for Majorana neutrinos (we discuss this case in Sec.~\ref{sec:FtonuR}). 

The Higgs portal contribution depends on the coupling of the SM Higgs boson $h$ to a pair of scalars $S$ after electroweak symmetry breaking. In the case of a complex scalar $S$, we parameterize it in terms of 
\begin{equation}
	 \mathcal{V}(S,h)\;\supset \; \lambda_{HS}\,v\, h (S^\dagger S) \label{Vb}
\end{equation}
and similarly for a real scalar $S$ with an additional factor $1/2$ in order for the Feynman rule (and therefore the expression of the Wilson coefficients) to be identical
\begin{equation}
	 \mathcal{V}(S,h)\;\supset \; \frac{\lambda_{HS}v}{2}\, h S^2\,. \label{VbrS}
\end{equation}
In the case of a complex scalar $S$, the Higgs couplings of Eq.~\eqref{Vb} are induced by SM gauge invariant Higgs portal interactions such as
\begin{align}
	(H^\dagger H) (S^\dagger S) &= h v S^\dagger S + \dots\label{eq:HHSS1}\\ 
	(H^\dagger S) (S^\dagger H) & = h v |S_d|^2 +\dots , \qquad\qquad \mathrm{with}\;S_d \equiv (S_{d,r}+i S_{d,i})/\sqrt{2}\,. \label{eq:HHSS2}\\
(H^\dagger S)^2+\,\text{H.c.} & =  hv \left(S_{d,r}^2  -S_{d,i}^2\right) +\dots \label{LNVquartic}\\
H^\dagger [\mathbf{S}^\dagger, \mathbf{S}] H& = h v (|S^+|^2-|S^-|^2)+\dots 
\end{align}
The term in the first line is always present, while those in the second and third
lines require $S$ to be an SU(2)$_\text{L}$ doublet, $S \equiv (S_u,S_d)^T$. Moreover the term in Eq.~\eqref{LNVquartic} assumes that $S$ has the same hypercharge as the SM Higgs doublet, which we write after spontaneous electroweak symmetry breaking as $H\equiv (0,(h+v)/\sqrt{2})^T$.
Finally the term in the last line exists for electroweak triplets $\mathbf{S}\equiv S\cdot \sigma$, where $S^\pm$ denotes the coefficients of $\sigma^\pm$.\footnote{$\sigma\equiv(\sigma_1,\sigma_2,\sigma_3)$ denote the Pauli matrices, with $\sigma^\pm=(\sigma_1 \pm i \sigma_2)/2$. } In the following we parameterize all the results in terms of $\lambda_{HS}$, which allows to easily
generalize the result of Higgs penguins for arbitrary combinations of Higgs portals. 
If $S^\pm$ ($S_{d,r}$ and
$S_{d,i}$) have the same mass, their contribution from the interactions \eqref{eq:HHSS2} and \eqref{LNVquartic} to the DD scattering amplitude exactly cancels due to the
relative minus sign in the interaction term.\footnote{This is not expected on
general grounds, as the same terms in the potential generate splittings after
electroweak symmetry breaking between the different components of the scalar
multiplets. Also a mass splitting, typically much smaller ($\mathcal{O}(100)$
MeV), is generated radiatively by loops of gauge bosons between the neutral and
the charged components of the SU(2)$_\text{L}$
multiplets~\cite{Cirelli:2007xd}.} For equal masses the effective coupling
$\lambda_{HS}$ can be generalized from the singlet case to an arbitrary
SU(2)$_\text{L}$ representation of dimension $d_F$ by replacing 
\begin{equation} \label{eq:LHSdF}
	\lambda_{HS} \to \begin{cases}
		2\, \lambda_{HS,1} + \lambda_{HS,2}  & \text{if $d_F=2$}\\
		d_F\, \lambda_{HS,1}  & \text{otherwise}
	\end{cases}
\end{equation}
where $\lambda_{HS,1}$ ($\lambda_{HS,2}$) is the coupling of the quartic scalar coupling in Eq.~\eqref{eq:HHSS1} (Eq.~\eqref{eq:HHSS2}).

\begin{table} [t!]
\centering
\begin{tabular}{l|c|ccc|c}
\toprule
Dark sector &Field & SU(3)$_{\rm C}$ & SU(2)$_{\rm L}$ & U(1)$_{\rm Y}$ & $Z_2$\\
\midrule
Dark matter&$\psi$ & 1 & 1 & 0 & $-1$\\ 
\midrule
Dark scalar&$S$ & 1 & $d_F$ & $Y_F$ & $\pm1$\\ 
Dark fermion &$F$ & 1 & $d_F$ & $Y_F$ & $\mp1$\\ 
\bottomrule
\end{tabular}
\caption{Particle content and quantum numbers of the Majorana DM scenario with a dark fermion and a dark scalar. The dark sector is charged under a $Z_2$ symmetry which stabilizes the DM $\psi$. }\label{tab:dark_sector_Majorana}
\end{table}

\subsection{Majorana dark matter} \label{sec:Maj}

If the DM particle is in a real representation of a stabilizing dark sector group, it could be a Majorana particle $\psi\equiv \psi_L+(\psi_L)^c$ (keeping the 4-component notation). We consider the simplest case of a $Z_2$ symmetry in the dark sector and comment on the general case in App.~\ref{ap:darkmattergroup}. The particle content for  Majorana DM  is listed in Tab.~\ref{tab:dark_sector_Majorana}. 
The Lagrangian is given by
\begin{eqnarray}
\begin{split}
	\mathcal{L}_{\psi} = &\;\frac12  \overline \psi\, (i\,\slashed \partial -m_{\psi})\,\psi \,+\,  i\, \overline F\, \slashed D \,F \, -\, m_{F}\, \overline{F}\, F + \left( D_\mu S\right)^\dagger D^\mu S - \mathcal{V}(S,H) \\
			     &  -\, \Big( y_1 \, \overline{F_{\rm R}} \, S\,\psi\,+y_2 \, \overline{F_{\rm L}} \, S\,\psi\,+\,\text{H.c.}\Big)\,. \label{Lpsimmaj}
\end{split}
\end{eqnarray}
If additionally $Y_F=0$ and consequently $S$ and $F$ both transform according to a real representation, they can be chosen to be a real scalar and a Majorana fermion $F=F_R+(F_R)^c$, respectively, and the fermionic part of the Lagrangian simplifies to
\begin{equation}
\mathcal{L}_{\psi}\;=\;\frac12  \overline \psi\, (i\,\slashed \partial -m_{\psi})\,\psi\,+\,  \frac12 \, \overline F\, (i \slashed \partial -m_{F}) \,F \,-\, \Big( y \, \overline{F} \, S\,\psi\,+ \text{H.c.}\Big) \label{Lpsibmaj}
\end{equation}
with $y=y_1=y_2$.

\subsection{Standard Model particles in the loop} \label{sec:darktoSM}
It is also interesting to study the case where one of the particles in the loop is a SM state. As either the scalar $S$ or the fermion $F$ need to be charged under the dark symmetry, only one of them can be substituted by a SM field. 
We discuss in the following the cases of $S$ being the Higgs doublet $H$, and $F$ being the lepton doublet $L$, the right-handed charged lepton $e_R$ or a right-handed neutrino $\nu_R$. Interestingly, these types of leptophilic models have some very nice features: 
	(i) the absence of  charged stable particles;
	(ii) the possibility to generate the correct relic abundance by annihilations into leptons;
	(iii) an interplay with LFV and leptonic AMMs;
	(iv) the possible relation to lepton number violation (LNV) and neutrino masses;
	(v) other possible phenomenological signals at future lepton colliders, like MET searches.

\mathversion{bold}
\subsubsection{Left-handed lepton doublet} \label{sec:FtoL}
\mathversion{normal}

The quantum numbers of the remaining states are fixed by demanding that the fermion $F$ in the loop is the SM lepton doublet $L$, as can be seen in Tab.~\ref{tab:dark_sectorL}. Moreover $y_1=0$ in Eq.~\eqref{Lpsid} for Dirac DM (or eq.~\eqref{Lpsimmaj} for Majorana DM), because we are now considering only chiral left-handed (LH) fermions.
\begin{table} [t]
\centering
\begin{tabular}{l|c|ccc|c}
\toprule
Sector &Field & SU(3)$_{\rm C}$ & SU(2)$_{\rm L}$ & U(1)$_{\rm Y}$ & $U(1)_{\rm dm}$\\
\midrule
Dark matter&$\psi$ & 1 & 1 & 0 & 1 \\ 
\midrule
Dark scalar&$S$ & 1 & 2 & $-1/2$ & $-1$\\ 
SM lepton doublet &$L$ & 1 & 2 & $-1/2$ & 0 \\ 
\bottomrule
\end{tabular}
\caption{Particle content and quantum numbers of the dark fermion scenario with the SM lepton doublet $L$ and a dark scalar. }\label{tab:dark_sectorL}
\end{table}
The coupling of the DM to the lepton doublets can lead to new contributions to LFV processes as well as AMMs of leptons, which are induced by loop diagrams with the dark scalar and the DM in the loop. These pose strong constraints on the flavor structure of the Yukawa couplings. However, the flavor constraints can be easily circumvented if DM only couples to the tau lepton.

In general, for direct couplings to leptons, it is possible to assign lepton number either to the DM particle
$\psi$ or the scalar $S$. An example with Majorana fermion DM
$\psi$ and a discrete $Z_2$ symmetry ($S\to -S$, $\psi\to -\psi$) is the
well-known scotogenic model, proposed in Ref.~\cite{Ma:2006km} and extensively
studied, e.g., in Refs.~\cite{Hessler:2016kwm,Vicente:2014wga,Molinaro:2014lfa,Toma:2013zsa,Racker:2013lua,Schmidt:2012yg,Gelmini:2009xd,Sierra:2008wj,Suematsu:2009ww,Hambye:2009pw,Kubo:2006yx}. See also the recent review on radiative neutrino mass models~\cite{Cai:2017jrq}. In this case
lepton number is broken by the combination of the Majorana mass term of $\psi$
and  the operator in Eq.~\eqref{LNVquartic}. These interactions generate
neutrino masses and lepton mixing at one-loop order, which significantly constrain the parameter space of the model.
However, in general DD and neutrino masses \emph{decouple}, because the LNV coupling in the potential could
be made arbitrarily small without affecting DD. 
For fermionic DM, typically, either coannihilations~\cite{Griest:1990kh} or the freeze-in mechanism \cite{McDonald:2001vt,Hall:2009bx}
need to be invoked in order to be compatible with low energy constraints,
specially the limit stemming from non-observation of $\mu \rightarrow e
\gamma$. 

\mathversion{bold}
\subsubsection{Right-handed charged lepton} \label{sec:Ftoe}
\mathversion{normal}

If $F$ is the SM right-handed (RH) charged lepton $e_{\rm R}$, the quantum numbers are fixed as shown in Tab.~\ref{tab:dark_sectorE}. In this case $y_2=0$ in Eq.~\eqref{Lpsid} for Dirac DM (or eq.~\eqref{Lpsimmaj} for Majorana DM), because the fermions have RH chirality.
\begin{table} [t!]
\centering
\begin{tabular}{l|c|ccc|c}
\toprule
Sector &Field & SU(3)$_{\rm C}$ & SU(2)$_{\rm L}$ & U(1)$_{\rm Y}$ & U(1)$_{\rm dm}$ \\
\midrule
Dark matter&$\psi$ & 1 & 1 & 0 & $1$\\ 
\midrule
Dark scalar&$S$ & 1 & 1 & $-1$ & $-1$\\ 
RH charged lepton &$e_{\rm R}$ & 1 & 1 & $-1$ & 0\\ 
\bottomrule
\end{tabular}
\caption{Particle content and quantum numbers of the dark fermion scenario with the SM right-handed charged lepton $e_{\rm R}$ and a dark scalar.}\label{tab:dark_sectorE}
\end{table}
As in the previous case one should expect new contributions to lepton AMMs and LFV processes.
By demanding that the scalar singlet $S$ has lepton number $+1$, the total lepton number is conserved  at the renormalizable level (the term in Eq.~\eqref{LNVquartic}) is absent) and consequently no Majorana neutrino masses are induced.

\mathversion{bold}
\subsubsection{Right-handed neutrino} \label{sec:FtonuR}
\mathversion{normal}

Dark matter may also couple to right-handed neutrinos $\nu_{\rm R}$ with $y_2=0$ in Eq.~\eqref{Lpsid} for Dirac DM (or eq.~\eqref{Lpsimmaj} for Majorana DM). In this case the quantum numbers are fixed as shown in Tab.~\ref{tab:dark_sectorN}.
\begin{table} [h]
\centering
\begin{tabular}{l|c|ccc|c}
\toprule
Sector &Field & SU(3)$_{\rm C}$ & SU(2)$_{\rm L}$ & U(1)$_{\rm Y}$ & U(1)$_{\rm dm}$ \\
\midrule
Dark matter&$\psi$ & 1 & 1 & 0 & $1$\\ 
\midrule
Dark scalar&$S$ & 1 & 1 & $0$ & $-1$\\ 
RH neutrino &$\nu_{\rm R}$ & 1 & 1 & $0$ & 0\\ 
\bottomrule
\end{tabular}
\caption{Particle content and quantum numbers of the dark fermion scenario with the SM right-handed  neutrino $\nu_{\rm R}$ and a dark scalar. }\label{tab:dark_sectorN}
\end{table}
As all particles in the loop are neutral, the only possible interactions are
with the $Z$ and Higgs bosons via the mixing of left- and right-handed neutrinos. This mixing is induced after electroweak symmetry breaking by 
\begin{equation}
\mathcal{L}_{\nu_{\rm R}}\;=\;
	-\overline L\, Y_\nu\, \nu_{\rm R}\,\tilde{H}\, - \frac{1}{2}\, \overline \nu_{\rm R} M_{\rm R} \nu_{\rm R}^c\,+{\rm H.c.}\,.\label{LpsidSS}
\end{equation}
In this scenario there are two possibilities regarding the nature of neutrinos: they are
Dirac fermions for $M_{\rm R}=0$, or Majorana fermions for $M_{\rm R} \neq 0$. 
In the latter case, Majorana masses for the
active light neutrinos are generated via the seesaw mechanism. In the seesaw
scenario the active-sterile mixing angles are tiny, either due to small Yukawa
couplings or large right-handed Majorana neutrino masses, and thus the $Z$
penguin contributions and the additional Higgs penguin contributions are
extremely small, which agrees with Eq.~(19) of
Ref.~\cite{Gonzalez-Macias:2016vxy}. A possible way-out is to consider an
inverse-seesaw scenario, where the suppression needed to have small neutrino
masses originates from a small LNV Majorana mass term, and not from small Yukawa couplings and/or large right-handed Majorana masses.

As DM couples to the SM particles mainly via neutrinos, this is known as the neutrino portal. It has been studied in detail for general heavy SM singlet Dirac and Majorana fermions $\nu_R$ in Ref.~\cite{Gonzalez-Macias:2016vxy} and also in Refs.~\cite{Escudero:2016ksa,Escudero:2016tzx}. 

\mathversion{bold}
\subsubsection{Higgs doublet} \label{sec:StoH}
\mathversion{normal}

Finally we consider the case of $S$ being the SM Higgs. This fixes the SM quantum numbers of the new particles, which are shown in Tab.~\ref{tab:dark_sectorH}.
\begin{table} [t]
\centering
\begin{tabular}{l|c|ccc|c}
\toprule
Sector &Field & SU(3)$_{\rm C}$ & SU(2)$_{\rm L}$ & U(1)$_{\rm Y}$ & U(1)$_{\rm dm}$ \\
\midrule
Dark matter&$\psi$ & 1 & 1 & 0 & 1\\ 
\midrule
SM Higgs doublet &$H$ & 1 & 2 & $1/2$ & 0\\ 
Dark fermion &$F$ & 1 & 2 & $1/2$ & 1 \\ 
\bottomrule
\end{tabular}
\caption{Particle content and quantum numbers of the DM fermion scenario with the SM Higgs and a dark fermion.}\label{tab:dark_sectorH}
\end{table}
This case is qualitatively different, because the neutral component of the
electroweak doublet $F$ and the fermion field $\psi$ mix after electroweak
symmetry breaking. The lighter of the two neutral mass eigenstates is the
DM particle. The Yukawa interactions with the Higgs necessarily induces
tree-level contributions to DD via Higgs and $Z$ boson exchange. Although a
tree-level contribution exists, DD may still be dominated by the loop-level
induced electric or magnetic dipole moments, because they are long-range
interactions.

\section{Effective operators for dark matter direct detection} \label{sec:opseft}

In the following sections we briefly review the effective operators for DM DD. In Sec.~\ref{sec:WCquarks} we show those involving DM interactions with quarks, while in Sec.~\ref{sec:WCnucleons} we briefly discuss their NR versions at the nucleon level. 

\subsection{Wilson coefficients at the quark level}  \label{sec:WCquarks}
Here we review the necessary notation for the effective interactions of the DM
with the quarks. The effective Lagrangian at the quark level for a DM fermion
$\psi$ is\footnote{We do not include twist-2 operators involving quarks and gluons in the effective Lagrangian. These are only generated by box diagrams, which are absent in our simplified models. They are relevant for example for wino DM in supersymmetric theories, see e.g.~Refs.~\cite{Drees:1993bu,Hisano:2015rsa}.}
\begin{equation}\label{Leff}
	\mathcal{L}_{\rm eff} \; = \; \sum_{k,q}\, c_k^{q} \,\mathcal{O}_k^{q} 
	+ c_g\, \mathcal{O}_g 
	+ \tilde c_g\, \tilde{\mathcal{O}}_g
	+ \mu_\psi \mathcal{O}_\text{mag}
	+ d_\psi \mathcal{O}_\text{edm}
	\,,
\end{equation}
where $c_k^{q}$ are the dimensionful Wilson coefficients with the quark $q$, $c_g$ and $\tilde c_g$ are the Wilson coefficients for gluon operators and $\mu_\psi$ and $d_\psi$ magnetic and electric dipole moments.
We implicitly assume that the operators are generated at a scale above the
nuclear scale, $\sim 2$ GeV. See App.~\ref{ap:NRmatching} for further details. 

We focus on the contributions to spin-independent (SI) and
spin-dependent (SD) operators of photon, $Z$ boson and Higgs penguins which are not momentum or velocity suppressed. The latter would yield very
small rates, as there is already the one-loop squared factor at cross section
level, $1/(16\pi^2)^2$. We start the discussion with the case of $\psi$
being a Dirac fermion and later on discuss the case of DM being a Majorana particle.

For SI scattering the relevant dimension-6 effective operators are
\begin{equation}
	\mathcal{O}^q_{\text{SS}} = m_q (\overline \psi \psi)  (\overline q q),\qquad \qquad
	\mathcal{O}^q_{\text{VV}} = (\overline \psi \gamma^\mu \psi)  (\overline q \gamma_\mu q)\,,
\end{equation}
where $q$ denotes the quark field. $\mathcal{O}^q_{\text{SS}}$ is generated by the gauge-invariant dimension-7 operators $(\overline \psi \psi)  (\overline Q_L \tilde H u_R)$ and $(\overline \psi \psi)  (\overline Q_L H d_R)$, where $Q_L,\,u_R,\,d_R$ represent the quark flavor eigenstates.
$\mathcal{O}^q_{\text{SS}}$ flips chirality and it is generated by Higgs
exchange and thus we factor out the quark mass $m_q$.
$\mathcal{O}^q_{\text{VV}}$ preserves chirality and is generated by photon or
$Z$ exchange.
The contribution from the photon penguin can be related to the anapole moment $\overline \psi \gamma^\mu \psi\,
\partial^\nu F_{\mu\nu}$ and the (non-gauge invariant) milli-charge
operator $\overline \psi \gamma^\mu \psi\,A_\mu$ via the equations of motion for the photon. 

There are also scatterings of the DM with gluons at two-loop order which generate the dimension-7 operators:
\begin{equation} \label{eq:gg}
	\mathcal{O}_{g}= \frac{\alpha_s}{12 \pi}(\overline \psi \psi)  G^{a\mu\nu} G^a_{\mu\nu},\qquad \qquad \tilde{\mathcal{O}}_{g} = \frac{\alpha_s}{8 \pi}(\overline \psi \psi)  G^{a\mu\nu} \tilde G^a_{\mu\nu}\,,
\end{equation}
where $a=1,...,8$ are the adjoint color indices, $\alpha_s$ is the strong coupling constant, $G_{\mu\nu}$ the gluon field strength tensor and $\tilde G_{\mu\nu} \equiv \tfrac12 \epsilon_{\mu \nu \rho \sigma} G^{\rho \sigma}$ its dual. $\mathcal{O}_g$ is induced from $\mathcal{O}_{SS}^q$ after integrating out the heavy quarks. We explicitly factorized out a loop factor, as these operators can never be generated at tree level.

For SD interactions the relevant dimension-6 effective operators are 
\begin{equation}
	\mathcal{O}_{\text{AA}}^q = (\overline \psi \gamma^\mu \gamma_5 \psi)  (\overline q \gamma_\mu \gamma_5 q), \qquad \qquad
	\mathcal{O}^q_{\rm TT} = (\overline \psi \sigma^{\mu \nu} \psi)  (\overline q \sigma_{\mu \nu}  q)\,,
\end{equation}
where $\sigma_{\mu\nu}=\tfrac{i}{2} [\gamma_\mu,\gamma_\nu]$. Only the $Z$
boson contributes to $\mathcal{O}^q_{\text{AA}} $. In SM effective
theory the tensor operator may arise from one of the dimension-7 operators
$(\overline \psi \sigma^{\mu\nu} \psi) (\overline Q_L \tilde H \sigma_{\mu\nu}
u_R)$ and $(\overline \psi \sigma^{\mu\nu}\psi) (\overline Q_L H
\sigma_{\mu\nu} d_R)$ which are however not induced at leading order. 

Photon penguins also generate long-range interactions which are described by the
magnetic (CP-even) and electric (CP-odd) dipole moments of the DM $\psi$, namely 
\begin{equation} \label{Leffmag}
\mathcal{O}_{\rm mag} = \frac{e}{8 \pi^2}(\overline \psi \sigma^{\mu\nu} \psi) F_{\mu\nu}, \qquad \qquad \mathcal{O}_{\rm edm} = \frac{e}{8 \pi^2}(\overline \psi \sigma^{\mu\nu} i \gamma_5 \psi) F_{\mu\nu}\,,
\end{equation}
with $\mu_\psi$ and $d_\psi$ the coefficients of the magnetic and electric dipole moment operators introduced in Eq.~\eqref{Leff}, respectively. The latter are generated radiatively and therefore it is convenient to factorize a loop factor.

In the case of a Majorana DM particle there are only operators with the bilinears $\overline{\psi}\psi$, $\overline{\psi}\,\gamma_5\psi$ and $\overline{\psi}\,\gamma^\mu\gamma_5\psi$, so that the vector $\mathcal{O}^q_{\rm VV}$, the tensor $\mathcal{O}^q_{\rm TT}$ and the dipole moment operators, $\mathcal{O}_{\rm mag}$ and $\mathcal{O}_{\rm edm}$, vanish identically. Thus, for SI scattering only the Higgs penguin which generates $\mathcal{O}^q_{\rm SS}$ is present. For SD scattering $\mathcal{O}^q_{\rm AA}$ generated by the $Z$ boson can also be non-vanishing. In this case we also compute the photonic contribution to the anapole operator 
\begin{equation} \label{Anapole}
	\mathcal{O}^q_{\rm AV} = (\overline \psi \gamma^\mu \gamma_5 \psi)  (\overline q \gamma_\mu q)\,,
\end{equation}
which gives rise to momentum-suppressed and velocity-suppressed NR operators (both SI and SD). See also Ref.~\cite{Matsumoto:2014rxa} for a study of the phenomenology of Majorana DM in EFT.

In general the penguin contributions are isospin-violating, i.e., with different couplings to protons and neutrons ($f_n \neq f_p$). This isospin violation is maximal for photon contributions which only couple to protons. The latter dominate the DM-nucleus scattering via the dipole moments $\mu_\psi$ and $d_\psi$. Hence for SI DM-nucleus scattering the enhancement due to coherent scattering is $Z^2$ instead of $A^2$, with $Z$ ($A$) being the number of protons (nucleons) of the nucleus.

\subsection{Non-relativistic Wilson coefficients at the nucleon level} \label{sec:WCnucleons}

The previous Wilson coefficients at the quark level generate non-trivial Wilson
coefficients at the nucleon level~\cite{Fan:2010gt,Fitzpatrick:2012ix,Fitzpatrick:2012ib}. The different contributions
generally interfere. 
The matrix elements of DM-nucleon scattering can be written as a linear combination of the following relevant NR operators
\begin{align}
	\mathcal{O}_1^N & = I_\psi I_N &
	\mathcal{O}_4^N & = \vec S_\psi \cdot \vec S_N \\
	\mathcal{O}_5^N & = \vec S_\psi\cdot \left(\vec v_\perp \times \frac{i\vec q}{m_N}\right)I_N & 
	\mathcal{O}_6^N & = \left(\vec S_\psi\cdot\frac{\vec q}{m_N}\right) \left(\vec S_N \cdot \frac{\vec q}{m_N}\right) \\
	\mathcal{O}_8^N & = \left(\vec S_\psi\cdot \vec v_\perp \right) I_N &
	\mathcal{O}_9^N & = \vec S_\psi \cdot \left(\frac{i\vec q}{m_N}\times \vec S_N\right) \\
\mathcal{O}_{11}^N & = - \left(\vec S_\psi \cdot \frac{i\vec q}{m_N}\right) I_N
\end{align}
in the convention of Ref.~\cite{Bishara:2017nnn}.
$I_\psi$ ($I_N$) denotes the identity operators for DM (nucleons), $\vec
S_\psi$ ($\vec S_N$) denotes DM (nucleon) spin, and $\vec q$ and $\vec v_\perp$
describe the momentum and velocity exchange. We use \texttt{DirectDM}~\cite{Bishara:2017nnn} to match the simplified models onto the NR operators. The numerical expressions for the matching to
NR operators are collected in App.~\ref{ap:NRmatching}. The NR Wilson coefficients may depend on the transferred
momentum $\vec q$. Note the different normalizations of the spinors and the
effective operators between
Refs.~\cite{Fan:2010gt,Fitzpatrick:2012ix,Fitzpatrick:2012ib,DelNobile:2013sia}
and Refs.~\cite{Anand:2013yka,Bishara:2016hek,Bishara:2017pfq,Bishara:2017nnn}.
In addition to the different definitions of the quark- and nucleon-level
operators, in order to translate between these conventions one needs to multiply
the NR Wilson coefficients of Refs.~\cite{Anand:2013yka,Bishara:2016hek,Bishara:2017pfq,Bishara:2017nnn} by $4\,m_\psi m_N$ ($4\,m_\psi |\vec q|^2$) in the case of contact (long-range) interactions. Further details can be found in the recent Refs.~\cite{DelNobile:2013sia,Anand:2013yka,Bishara:2016hek,Bishara:2017pfq,Bishara:2017nnn}. The differential cross section for DM scattering off nuclei is given in App.~\ref{ap:DD}.

\section{Analytical results} \label{sec:computation}
The effective operators in Eq.~\eqref{Leff} are generated at one-loop order
from penguin diagrams mediated by the photon and the $Z$ and Higgs bosons.  We
have computed the different contributions using the Mathematica packages
\texttt{FeynRules}~\cite{Alloul:2013bka}, \texttt{FeynArts}~\cite{Hahn:2000kx},
\texttt{FormCalc} and
\texttt{LoopTools}~\cite{Hahn:1998yk,Christensen:2009jx,Hahn:2006zy},
\texttt{ANT}~\cite{Angel:2013hla} and \texttt{Package
X}~\cite{Patel:2015tea,Patel:2016fam}. As we show below, although the long-range
interactions are expected to dominate, the short-range effective operators
become relevant in some cases. One obvious example is DM-nucleus scattering of
Majorana DM, since the dipole moments vanish. Therefore we show below all
relevant contributions.

The interesting SI (SD) interactions in Eq.~\eqref{Leff} are given by the dipole moment operators $\mathcal{O}_\text{mag}$ and $\mathcal{O}_\text{edm}$ as well as the operators  $\mathcal{O}^q_{\rm SS}$, $\mathcal{O}_g$ and $\mathcal{O}^q_{\rm VV}$ ($\mathcal{O}^q_{\rm AA}$). All the other operators in Eq.~\eqref{Leff} are suppressed in the limit of  small momentum transfer by a factor $|\vec q|^2/m_N^2$ or $|\vec q|^2/m_\psi^2$, where $m_N$ is the nucleon mass. 
In the following we express the SI and SD Wilson coefficients in Eq.~\eqref{Leff} in terms of the ratios
\begin{equation}
 x_\psi \; \equiv \; \frac{m_\psi}{m_S} 	\quad\quad\text{and} \quad\quad  x_F \; = \;  \frac{m_F}{m_S}\,,
 \end{equation}
and the loop function 
\begin{equation}
	g\big(x_\psi,x_F\big) \; = \; \frac{\ln\Bigg(\frac{1-x_\psi^2+x_F^2+\sqrt{x_\psi^4+(1-x_F^2)^2-2x_\psi^2(1+x_F^2)}}{2 x_F}\Bigg)}{\sqrt{x_\psi^4+(1-x_F^2)^2-2x_\psi^2(1+x_F^2)}}\,.
\end{equation}
It is convenient to define the vector and axial Yukawa couplings:
\begin{equation} \label{eq:yVyA}
	y_V \; \equiv \; \frac 1 2\left(y_1+y_2 \right)\,,\quad\quad y_A \; \equiv \; \frac 1 2\left(y_2-y_1 \right)\,.
\end{equation}
Similarly, the interaction of the fermion $F$ with the $Z$ boson in Eq.~\eqref{Lpsid} may be written in terms of vector and axial-vector couplings, namely
\begin{equation}
	\mathcal{L}_F^Z \; =\;  \frac{e}{c_w\, s_w}\,Z^\mu\,\overline{F}\,\gamma_\mu\left(z_V - z_A\,\gamma_5\right)F \,,
\end{equation}
where $e>0$ is the proton electric charge, and $s_w\,(c_w)$ denotes the sine (cosine) of the weak mixing angle.
If $F$ is a vector-like fermion, then we have 
\begin{equation}
	z_V \; = \; c_w^2\,Q - Y_F\,,\quad\quad\quad z_A\;=\;0\,,
\end{equation}
where $Q$ is the electric charge of the (component of the) field $F$, in units of $e$,  and $Y_F$ is the corresponding hypercharge.
Conversely, for a SM lepton $F$ we have
\begin{equation}\label{ZFSM}
	z_V \; = \; \frac 12\left( (1-2 s_w^2)\,Q - Y_F\right)\,,\quad\quad\quad z_A\;=\; \frac 12 \left(Q-Y_F \right)\,,
\end{equation}
and the Yukawa couplings are
\begin{eqnarray}
\begin{split}
	y_V & = \;  y_A \; = \; \frac{y_2}{2} \quad\quad \quad    \,\;\;\; \text{if $S$ is a doublet of SU(2)$_\text{L}$}\,,\\
	y_V & = \;  -\,y_A \; = \; \frac{y_1}{2} \quad\quad\quad     \, \text{if $S$ is a singlet of SU(2)$_\text{L}$}\,.
\end{split}	
\end{eqnarray}
For simplicity of notation we report the full analytic results for SU(2)$_{\rm L}$ singlets $F$ and $S$. In the case of no mass splittings between the components of the SU(2)$_{\rm L}$ multiplets of dimension $d_F$ it is straightforward to generalize the results:
The expressions for photon penguins and electric and magnetic dipole moments are generalized by replacing $Q\to d_F\,Y_F$. 
Higgs penguins are generalized for different scalar multiplets as in Eq.~\eqref{eq:LHSdF}.

Most $Z$ penguin contributions (apart from some with chiral SM fermions) vanish at leading order. This is also the case for other SU(2)$_{\rm L}$ multiplets.
 
We summarize below the relevant contributions to the (Dirac or Majorana)
DM--quark scattering amplitude. We have checked that our expressions agree with
those reported in the literature in the appropriate limits: dipole and anapole
moments in Refs.~\cite{Agrawal:2011ze,Kopp:2014tsa,
Schmidt:2012yg,Ibarra:2015fqa}, and also for the $Z$ boson contributions in
Refs.~\cite{Berlin:2015njh,Ibarra:2015fqa}.

\subsection{Dirac dark matter}
The leading contributions for Dirac fermion DM are from dipole moments, the operators $\mathcal{O}_{VV}^q$ and $\mathcal{O}_{AA}^q$, and the scalar operator $\mathcal{O}_{SS}^q$. Integrating out heavy quarks induces the gluon operator $\mathcal{O}_g$.

\subsubsection{Electromagnetic dipole moments} 

The magnetic and electric dipole moments are given by 
\begin{eqnarray}
\begin{split}
	 \mu_\psi  = &\; -\frac{Q}{4\,x_\psi^3\, m_S} \,\left |y_V\right |^2 \Bigg[ x_\psi^2+\left(1-x_\psi x_F - x_F^2 \right)\ln x_F \\
	&\; -\, \Big(x_\psi^3 x_F-(1-x_F^2)^2+x_\psi^2(1+x_F^2)+x_\psi x_F(1-x_F^2)\Big)g\big(x_\psi, x_F\big)\Bigg]\\
	&\; -\, \big(y_V\to y_A,\,x_\psi\to-x_\psi,\,x_F\to x_F\big)\,,
\end{split}
\end{eqnarray}
and
\begin{eqnarray}
\begin{split}
	d_\psi  = & \; -\frac{Q}{2\,x_\psi^2\, m_S} \,\text{Im}[y_V\,y_A^*]\,x_F\Bigg[\ln x_F + \Big(1+x_\psi^2-x_F^2\Big) g\big(x_\psi, x_F\big)\Bigg] \,.
\end{split}
\end{eqnarray}
Both $\mathcal{O}_{\rm mag} $ and $\mathcal{O}_{\rm edm}$ flip chirality and
therefore the dominant contributions to their coefficients are proportional
to the heaviest fermion mass, either $m_\psi$ or $m_F$.
In the limit $m_\psi \ll m_F<m_S$ these expressions reduce to
\begin{eqnarray}
	\mu_\psi & \approx & -\frac{Q}{4\, m_S}\left(\left |y_V\right |^2 -\left |y_A\right |^2 \right)x_F \,\frac{1-x_F^2+2\,\ln x_F}{(1-x_F^2)^2}\nonumber \\
	 && + \frac{Q}{8\, m_S}\left(\left |y_V\right |^2+\left |y_A\right |^2 \right)x_\psi \frac{1-x_F^2(x_F^2-4\ln x_F)}{(1-x_F^2)^3}\,,\\
	d_\psi & \approx &	 -\frac{Q}{2\, m_S} \,\text{Im}[y_V^*\,y_A]\,x_F \,\frac{1-x_F^2+2\,\ln x_F}{(1-x_F^2)^2}	\,.
\end{eqnarray}

\subsubsection{Photon penguin} 

Photon penguins induce the operator $\mathcal{O}_{VV}^q$. The relevant Wilson coefficient in the effective Lagrangian (\ref{Leff}) is
\begin{eqnarray}
\begin{split}
	c^q_{\rm VV}  = &\; - \frac{\alpha_{\rm em}}{24\,\pi\,x_\psi^ 4}\,\frac{1}{m_S^2}\,Q\,Q_q\,\left |y_V\right |^2\\
	&\; \Bigg[ \Big(-3x_\psi^6+6x_\psi^5x_F+12x_\psi x_F(1-x_F^2)^2+8(1-x_F^2)^3 +2x_\psi^4(5+x_F^2) \\
	& \;-6x_\psi^3x_F(1+3x_F^2)
	-3x_\psi^2(5-2x_F^2-3x_F^4)\Big)\frac{g\big(x_\psi, x_F\big)}
	{1-(x_\psi-x_F)^2}\\
	&\;+\frac{2 x_\psi^2(4-3 x_\psi^2+6 x_\psi x_F-4 x_F^2)}{1-(x_\psi-x_F)^2} +(8+x_\psi^2-4 x_\psi  x_F-8 x_F^2) \ln x_F\Bigg]\\
	&\;+ \big(y_V\to y_A,\,x_\psi\to-x_\psi,\,x_F\to x_F\big)\,,
\end{split}	
\end{eqnarray}
where $Q_q$  is  the electric charge of the quark $q$  in units of $e>0$.  
In the limit $m_\psi \ll m_F<m_S$ the expression above reduces to 
\begin{eqnarray}
	c^q_{\rm VV}  \;\approx \;\frac{Q \,Q_q \,\alpha_{\rm em}}{24\,\pi\,m_S^2}\,\big(\left|y_V\right|^2+\left|y_A\right|^2\big)\,
	 \frac{3-3\,x_F^2+2\,(2+x_F^2)\ln x_F}{(1-x_F^2)^2}\,.
\end{eqnarray}
In the case the mass of the fermion in the loop is much smaller than the momentum transfer, $m_F\ll\sqrt{-q^2}$, we have
\begin{eqnarray}
\begin{split}
	c^q_{\rm VV}   \approx & \;\frac{Q \,Q_q \,\alpha_{\rm em}}{72\,\pi\,m_S^2}\,\big(\left|y_V\right|^2+\left|y_A\right|^2\big)\\\,
	                    & \;  \Bigg[ \frac{12 x_\psi^4\ln x_q -8 x_\psi^2 (3-x_\psi^2)-3(8-7x_\psi^2+3x_\psi^4)\ln\big(1-x_\psi^2\big)}{x_\psi^4(1-x_\psi^2)}\Bigg] 
\end{split}	                    
\end{eqnarray}
with $x_q\equiv \sqrt{-q^2/m_S^2}$.

\mathversion{bold}
\subsubsection{$Z$ penguin}\label{sec:ZDirac}
\mathversion{normal}
 
For a vector-like fermion the resulting SI and SD scattering amplitudes are suppressed by $|\vec q|^2/m_F^2$ and $|\vec q|^2/m_S^2$ due to a cancellation between the diagrams where the $Z$ boson couples to the scalar and to the fermion. Therefore, no strong constraints on the model parameters can be obtained. For SM leptons in the loop we distinguish two cases: 

(i) If $S$ is a singlet under SU(2)$_{\rm L}$, the axial-vector coupling in Eq.~\eqref{ZFSM} is $z_A=0$ and both SI and SD scattering amplitudes are suppressed as for a vector-like fermion.

(ii) If $S\equiv(S^0,S^-)^T$ is a doublet under SU(2)$_{\rm L}$, there are contributions from both diagrams where the $Z$ boson is attached to the SM lepton or the scalar in the loop
\begin{eqnarray}
	c^q_{\rm VV} & =& \sum_{f=\{\ell,\nu\}} \frac{ (1+2\, Q_f)\alpha_{\rm em}}{16\,\pi\,c_w\,s_w\,m_Z^2} \frac{q_V}{e} 
	\,\frac{x_f^2}{x_\psi^2}\,\left| y_2\right|^2 \bigg[ \left(x_f^2 - 1 - x_\psi^2 \right)g\left(x_\psi, x_f \right)- \ln x_f \bigg]\,,  \label{eq:cVVDirac}\\
	c^q_{\rm AA} & =& \; c^q_{\rm VV}\, (q_V \rightarrow q_A)\,. \label{eq:cAVDirac}
\end{eqnarray}
The couplings $q_{V,A}$ are $q_V/e = 3-8s_w^2/(12 c_w s_w)$ and $q_A/e = -1/(4 c_w
s_w)$ for up-type quarks and $q_V/e = -3+4s_w^2/(12 c_w s_w)$ and $q_A/e =
1/(4 c_w s_w)$ for down-type quarks. $Q_f$ denotes the electric charge of the
lepton. We define $x_\psi \equiv m_\psi/m_{S^-}$, $x_\ell \equiv
m_\ell/m_{S^-}$ and $x_\nu\equiv m_\nu/m_{S^0}$ with the charged lepton mass
$m_\ell$ and the neutrino mass $m_\nu$. This agrees with the expression in
Ref.~\cite{Ibarra:2015fqa}. The contribution with light active neutrinos in the
loop is negligible because it is proportional to $x_\nu^2$ and thus the
contribution is entirely determined by the charged lepton in the loop. However,
for models with a neutrino portal as outlined in Sec.~\ref{sec:FtonuR} there
may be a sizable contribution from right-handed neutrinos (mixed with left-handed neutrinos) in the loop. In the limit of small DM mass, $x_\psi\ll 1$, the
contribution of right-handed neutrinos is
\begin{align} \label{eq:cVVN}
	c_{{\rm VV},N}^q & = \frac{ \alpha_{\rm em}\, \sin^2\theta }{16\pi c_w s_w m_Z^2} \frac{q_V}{e} |y_2|^2 \frac{x_{N}^2}{(1-x_N^2)^2} \left(1-x_N^2 +2\ln x_N\right) \\
	c_{{\rm AA},N}^q & = c_{{\rm VV},N}^q(q_V\to q_A)  \label{eq:cAAN}
\end{align}
with the active-sterile mixing angle $\theta$. We define $x_{N}\equiv m_{N}/m_S$ with the heavy neutrino mass $m_N$.
These interactions are also discussed in Ref.~\cite{Gonzalez-Macias:2016vxy} (see also Ref.~\cite{Batell:2017cmf}). In the case of mixing of vector-like charged fermions with SM charged leptons, there is an overall minus sign in the expressions of Eqs.~\eqref{eq:cVVN} and~\eqref{eq:cAAN}.

\subsubsection{Higgs penguin}

At leading order in $|\vec q|^2$ there is only the contribution to the SI scattering amplitude.
The relevant Wilson coefficient generated by the Higgs-portal
interaction is 
\begin{eqnarray}
\begin{split} \label{eq:cHSI}
	c^q_{\rm SS}  = & \; - \frac{\lambda_{HS}}{16\, \pi^2\, x_\psi^3\,m_h^2} \,\frac{1}{m_S}\,\left |y_V\right |^2   \Bigg[ x_\psi^2 + \left( 1- x_\psi^2 - x_\psi x_F - x_F^2 \right)\ln x_F  \\
	& \; + \left(1- x_\psi^2 - 2x_\psi x_F - x_F^2  \right)   \left(1- x_\psi^2 + x_\psi x_F - x_F^2  \right) g\left( x_\psi, x_F\right) \Bigg]\\
	&\; -\, \big(y_V\to y_A,\,x_\psi\to-x_\psi,\,x_F\to x_F\big)\,.
\end{split}	
\end{eqnarray}
As previously mentioned we neglect the contribution from the Higgs penguin where the Higgs boson couples to a SM lepton in the loop, because it is suppressed by $(m_{\ell}/v)^2\ll 1$.\footnote {For the case of SM leptons in the loop this other
contribution of the Higgs coupling to the leptons is given in
Ref.~\cite{Ibarra:2015fqa}. For the neutrino portal these interactions are
given in Ref.~\cite{Gonzalez-Macias:2016vxy}.\label{foot}} 
As in the case of $\mathcal{O}_{\rm mag} $ and $\mathcal{O}_{\rm edm}$, the operator $\mathcal{O}_{SS}^q$ flips chirality, and therefore the dominant contribution to
its Wilson coefficients is proportional to either $m_\psi$ or $m_F$. If
both $F$ and $\psi$ are charged under U(1)$_{\rm dm}$, then $m_\psi< m_F$ and
thus the largest contribution comes with the chirality flip on the fermion line of $F$. On the contrary if $F$ is a SM lepton the largest Higgs contribution is proportional to $m_\psi$. In the limit $m_\psi \ll m_F<m_S$, Eq.~\eqref{eq:cHSI} simplifies to
\begin{eqnarray}
\begin{split}
	c^q_{\rm SS}  \approx & \; -\frac{\lambda_{HS}}{16\, \pi^2\, m_h^2\,m_S\,(x_F^2-1)^2} \,\Bigg\{x_F\,(\left |y_V\right |^2-\left |y_A\right |^2) \,  \left(-x_F^2+2 \,x_F^2\,\ln\left(x_F\right)+1\right)+\\
	&\; +\frac{x_\psi}{2\,(x_F^2-1)}\,(\left |y_V\right |^2+\left |y_A\right |^2) \, \left(-3\,x_F^4+4 \,x_F^4\,\ln\left(x_F\right)+4\,x_F^2-1\right) \Bigg\}\,.
\end{split}	
\end{eqnarray}

\subsection{Majorana dark matter}

\subsubsection{Photon penguin}

For Majorana DM $\psi$ the electromagnetic dipole moments identically vanish and the only allowed electromagnetic form factor is the anapole moment. This gives rise to the effective operator $\mathcal{O}^q_{\rm  AV}$ in Eq.~\eqref{Anapole}. We obtain
\begin{equation}
	c^q_{\rm AV}  \;= \; -\frac{Q \,Q_q \,\alpha_{\rm em}}{2\,\pi\,x_\psi^2\,m_S^2}\,\text{Re}[y_V^*\,y_A]\,\Bigg[\ln x_F + \left(1+\frac{x_\psi^2}{3}-x_F^2\right) g\left( x_\psi, x_F\right)\Bigg]\,.
\end{equation}
In the limit $m_\psi \ll m_F<m_S$ this simplifies to
\begin{equation}
c^q_{\rm AV}  \;\approx \;\frac{Q \,Q_q \,\alpha_{\rm em}}{6\,\pi\,m_S^2}\,\text{Re}[y_V^*\,y_A]\,\
	 \frac{3-3\,x_F^2+2\,(2+x_F^2)\ln x_F}{(1-x_F^2)^2}\,.
\end{equation}
In the case the mass of the fermion in the loop is much smaller than the momentum transfer, $m_F\ll\sqrt{-q^2}$, we have
\begin{eqnarray}
\begin{split}
	c^q_{\rm AV} = & \; -\frac{Q \,Q_q \,\alpha_{\rm em}}{18\,\pi\,m_S^2}\,\text{Re}[y_V^*\,y_A]\, \frac{2x_\psi^2 (5-6\ln x_q)+3(3+x_\psi^2)\ln\big(1-x_\psi^2\big) }{x_\psi^2(1-x_\psi^2)}\,,
\end{split}	
\end{eqnarray}
where $x_q\equiv \sqrt{-q^2/m_S^2}$.

\mathversion{bold}
\subsubsection{$Z$ and Higgs penguin} \label{sec:ZHMaj}
\mathversion{normal}

For a vector-like fermion $F$ in the loop the $Z$ penguin diagram does not contribute to the SI scattering amplitude, because the DM vector current identically vanishes for Majorana fermions. The SD scattering amplitude is suppressed by $|\vec q|^2/m_F^2$ and $|\vec q|^2/m_S^2$ due to a cancellation similar to that occurring in the case of Dirac fermion DM, see Sec.~\ref{sec:ZDirac}.
If $F$ is a left-handed lepton doublet, and consequently $S\equiv(S^0,S^-)^T$ is an SU(2)$_{\rm L}$ doublet, we find at leading order in $|\vec q|^2$: $c^q_{\rm VV}  = 0$ and $c^q_{\rm AA}$ is a factor of two larger than result for the Dirac case provided in Eq.~\eqref{eq:cAVDirac}.
If $F$ is a right-handed charged lepton or a right-handed
neutrino, the scalar $S$ is necessarily an SU(2)$_{\rm
L}$-singlet and thus the axial-vector coupling $z_A$ in
Eq.~\eqref{ZFSM} is zero and both SI and SD scattering
$Z$-mediated amplitudes are suppressed as for a vector-like
fermion. In some models, like with right-handed neutrinos or with vector-like fermions, there can be mixing with SM leptons. These generate couplings to the $Z$ and the Higgs bosons, see discussion around Eqs.~\eqref{eq:cVVN} and ~\eqref{eq:cAAN}, and footnote \ref{foot}.

For the Higgs penguin there is only a contribution to the SI amplitude $c^q_{\rm SS}$ at leading order in $|\vec q|^2$, which again is a factor of two larger than in the Dirac DM case, given in Eq.~\eqref{eq:cHSI}. The fact that the $h$ and the $Z$ penguin contributions to the non-zero Wilson coefficients, $c^q_{\rm SS}$ and $c^q_{\rm AA}$, are a factor of 2 larger for Majorana than for Dirac DM, can be understood from the presence of extra crossed diagrams for Majorana particles, where the initial and final DM particles are interchanged.

\section{Numerical analysis}  \label{sec:numerics}

We use \texttt{LikeDM}~\cite{Huang:2016pxg,Liu:2017kmx} to compute the differential rates
and the experimental upper bounds on our scenarios. We have also performed cross checks with the program of Ref.~\cite{DelNobile:2013sia}. First we show results
for the event rates and upper limits for Dirac and Majorana DM, having either
vector-like fermions or SM leptons in the loop. For the latter case we also show
upper limits from LFV signals. In the following we parameterize the vector and
axial Yukawa couplings of Eq.~\eqref{eq:yVyA} in terms of their absolute value and phase as $y_V=|y_V| e^{i \phi_V}$ and $y_A=|y_A| e^{i \phi_A}$.

\subsection{Wilson coefficients at the quark level} \label{sec:ex_quark}
\begin{figure}[t!]
	\centering
	\begin{subfigure}{\linewidth}
		\includegraphics[width=0.48\linewidth]{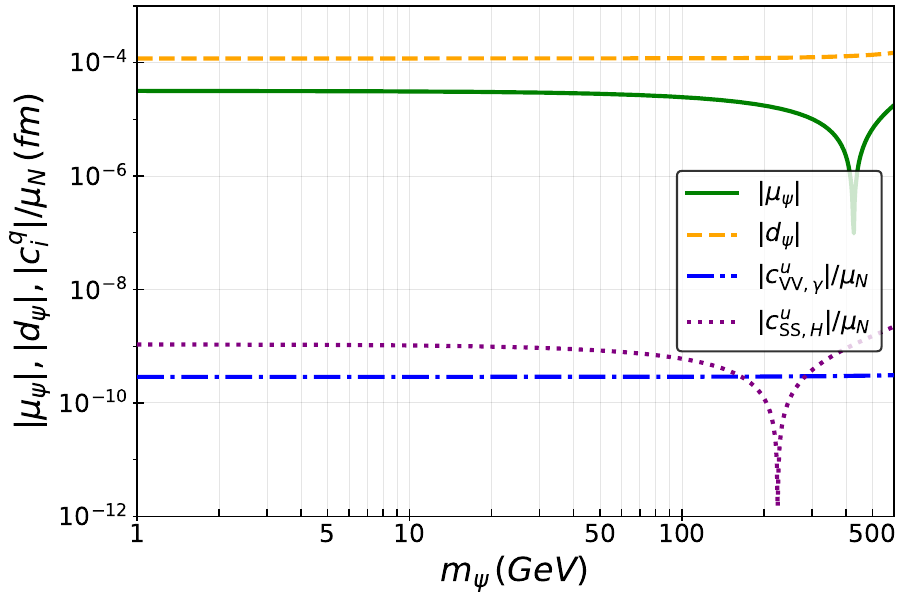}\hfill
		\includegraphics[width=0.48\linewidth]{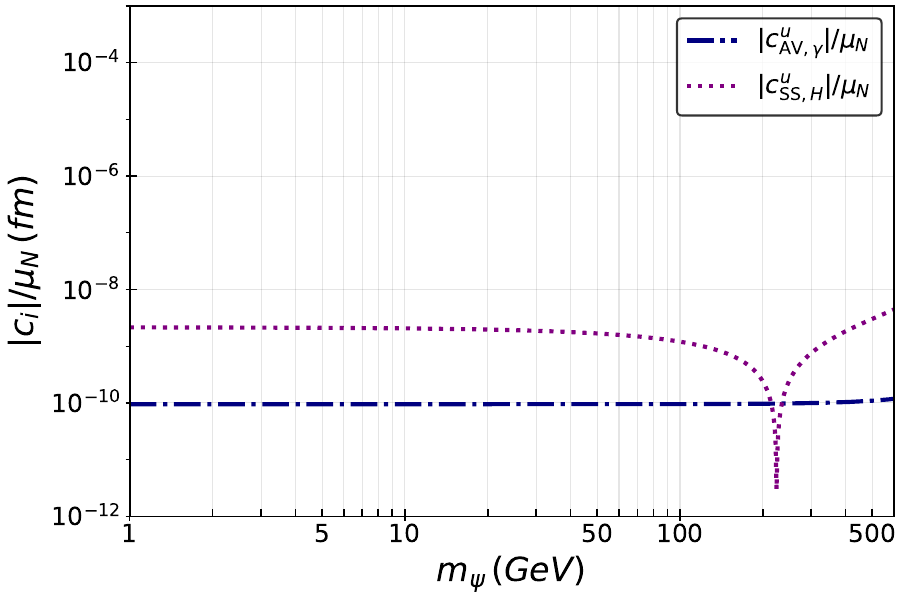}
		\caption{Dirac DM (left) and Majorana DM (right) with a vector-like fermion $F$ of mass $m_F=600$ GeV and a scalar $S$ of mass $m_S=500$ GeV. The Yukawa couplings are fixed to $y_V=1$ and $y_A=1.3\, e^{i\,1.4}$.} 
	\end{subfigure}

	\vspace{2ex}
	\begin{subfigure}{\linewidth}
		\includegraphics[width=0.48\linewidth]{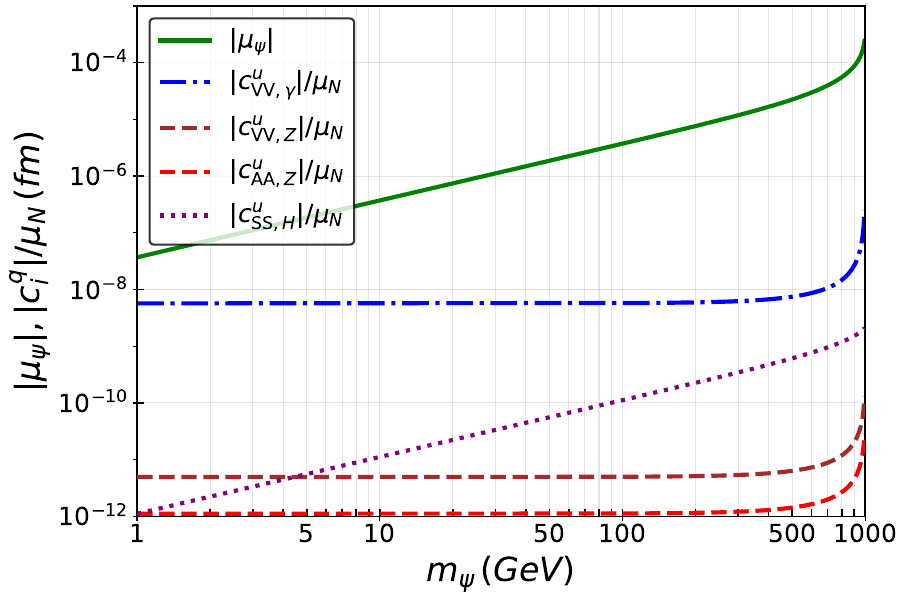}\hfill
		\includegraphics[width=0.48\linewidth]{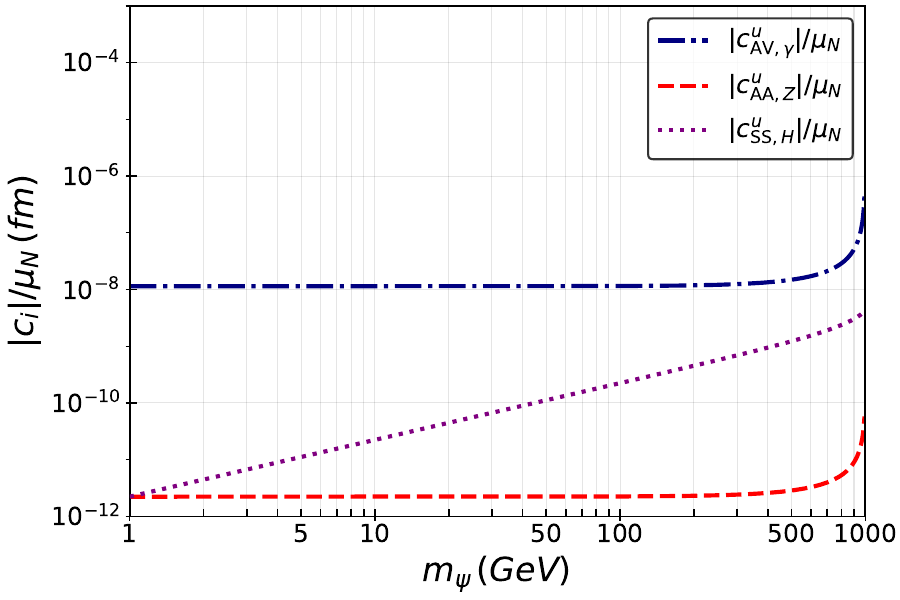}
		\caption{Dirac DM (left) and Majorana DM (right) with left-handed SM leptons in the loop and a scalar $S$ of mass $m_S=1000$ GeV. All Yukawa couplings are fixed to $y_2=1$.} 
	\end{subfigure}

	\caption{Wilson coefficients at the quark level (with up-type quarks) versus the DM mass $m_\psi$. The Higgs portal coupling is $\lambda_{HS}=3$. The vector and scalar coefficients which originate from the photon, $Z$ and Higgs penguin diagrams, respectively, have been rescaled by the nuclear magneton $\mu_N=e/(2m_p)$. 
	The photon penguin contribution $c^u_{{\rm VV},\gamma}$ depends on the transferred momentum $q$ for light SM leptons: We choose a recoil energy $E_R=8.59$ keV for ${}^{132}_{54}\mathrm{Xe}$ which results in $|\vec q|^2=2.11 \times 10^{-3}\, \mathrm{GeV}^2$.}
\label{fig:cont}
\end{figure}

In order to illustrate the relative weight of the different contributions, we
plot in Fig.~\ref{fig:cont} the long and short-range contributions with up-type
quarks for vector-like fermions (upper panel) and for a SM left-handed lepton
doublet (lower panel) in the loop. The plots on the left correspond to Dirac DM, while the plots on the right are for Majorana DM. Unless otherwise stated we always
set the dark charge $Q_\psi$ to one and fix the Higgs portal coupling,
$\lambda_{HS}=3$. The Wilson coefficients of the short-range interactions
(dimension-6 operators) have been rescaled by the nuclear magneton
$\mu_N=e/(2m_p)$ to compare them to the (dimension-5) dipole moments.

For Dirac DM with vector-like fermions in the loop (top left) we show the magnetic moment
$\mu_\psi$ (in solid green), the dipole moment $d_\psi$ (dashed orange), as
well as the short-range contributions mediated by the photon (dot-dashed
blue) and the Higgs (dotted purple). We have fixed $m_F=600$ GeV, $m_S=500$
GeV, $y_V=1$ and $y_A=1.3\, e^{1.4\,i}$. The DM electric
dipole moment $d_\psi$ is around $10^{-4}\, \mathrm{fm}$ and it dominates, followed closely by the magnetic moment. The Higgs and the short-range
photon interactions are always very suppressed, below $10^{-9} \, \mathrm{fm}$. All Wilson coefficients increase
for $m_\psi \sim m_F + m_S$ (not shown as we demand $\psi$ to be the lightest particle charged under U(1)$_{\rm dm}$), when the particles in the loop are almost on-shell. For this example the Wilson coefficients $\mu_\psi$ and $c^u_{\rm SS}$ change sign at particular values of the DM mass and thus there is a dip in their absolute magnitude. The case of Majorana DM with vector-like fermions (top right) only shows the short-range Higgs and photon contributions, the latter being the anapole
moment (dot-dashed dark blue). These Wilson coefficients are of similar size
as in the Dirac case, although the photon anapole (Higgs) contribution is
smaller (a factor of two larger) than the photon short-range (Higgs) Wilson
coefficient of the Dirac case.

For Dirac DM with SM lepton doublets in the loop (bottom left) we show the magnetic moment $\mu_\psi$ (in solid green), the short-range contributions mediated by the
photon (dot-dashed blue), the $Z$ penguin SI (dashed brown) and SD (dashed red) scattering and the Higgs penguin (dotted purple). The electric dipole moment $d_\psi$
vanishes at one-loop order. We have fixed $m_S=1000$ GeV and $y_V=y_A=y_2/2=1/2$. In the case of the (light) SM
leptons in the loop, the photon penguin contribution
$c^u_{{\rm VV},\gamma}$ depends on the transferred momentum $\sqrt{2 m_A E_R}$,\footnote{$m_A$ is the nucleus mass and $E_R$ the  recoil energy.} for which we use $E_R=8.59$ keV (which is a reasonable value for xenon
nuclei, with mass $m_{\rm Xe} \simeq 132$ GeV). The magnetic dipole moment dominates, followed by the
photon short-range contribution which is roughly $\sim 10^{-8}\, \mathrm{fm}$. The
increase of $\mu_\psi$ and the Higgs contribution with $m_\psi$ is easily
understood from chirality arguments. This also implies that $\mu_\psi$ and
$c^u_{\rm SS}$ are suppressed with respect to the case of vector-like leptons
(cf. upper-left panel of Fig.~\ref{fig:cont}) by the DM mass, except in the
region of $m_\psi$ close to $m_S$. The Higgs and the $Z$ penguin interactions are
always very suppressed (for the $Z$ penguin the SD amplitude is smaller than the SI
contribution, due to the factors $q_{V,A}/e$ in
Eqs.~\eqref{eq:cVVDirac} and ~\eqref{eq:cAVDirac}), below $10^{-11} \, \mathrm{fm}$, and therefore they can be  safely
neglected. All Wilson coefficients increase for $m_\psi \sim m_F +
m_S$.

For Majorana DM with SM lepton doublets in the loop (bottom right) the Higgs and the $Z$ SD amplitudes are a factor of two larger than in the Dirac case and with the same dependence on $m_\psi$, while the anapole Wilson coefficient (dot-dashed purple) is slightly larger than the photon short-range contribution present in the Dirac case. Notice that this is the opposite behavior of the case with vector-like fermions. 

\subsection{Wilson coefficients at the nucleon level}  \label{sec:ex_nucleon}

\begin{figure}[t!]
	\centering
	\begin{subfigure}{\linewidth}
		\includegraphics[width=0.48\linewidth]{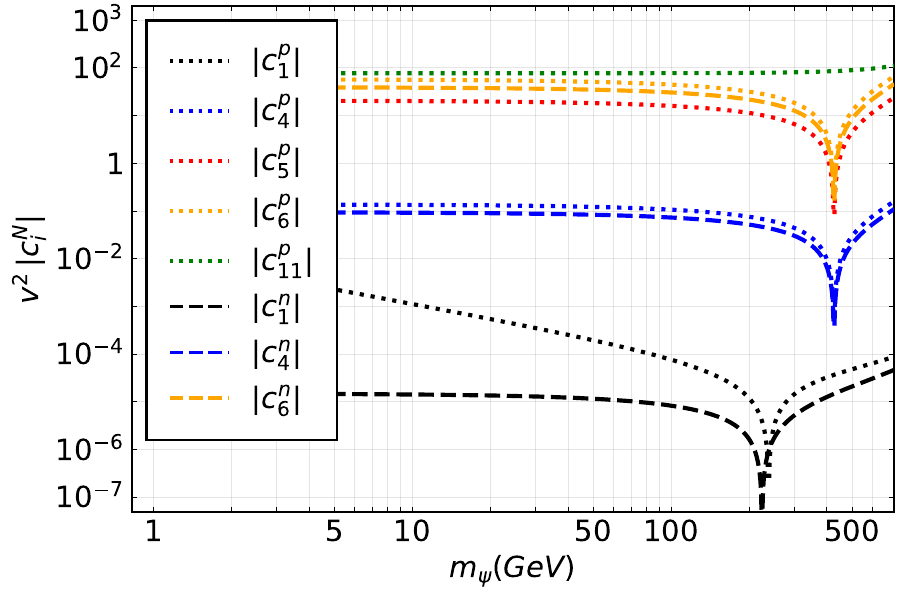}\hfill
		\includegraphics[width=0.48\linewidth]{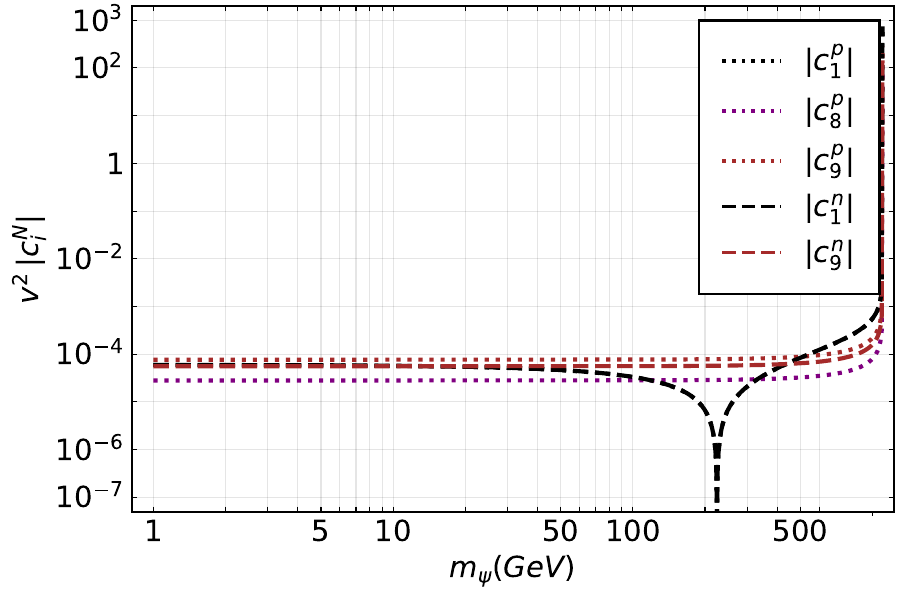}
		\caption{Dirac DM (left) and Majorana DM (right) with a vector-like fermion $F$ of mass $m_F=600$ GeV and a scalar $S$ of mass $m_S=500$ GeV. The Yukawa couplings are fixed to $y_V=1$ and $y_A=1.3\, e^{i\,1.4}$.} 
		\label{fig:NRVL}
	\end{subfigure}
	
	\vspace{3ex}
	\begin{subfigure}{\linewidth}
		\includegraphics[width=0.48\linewidth]{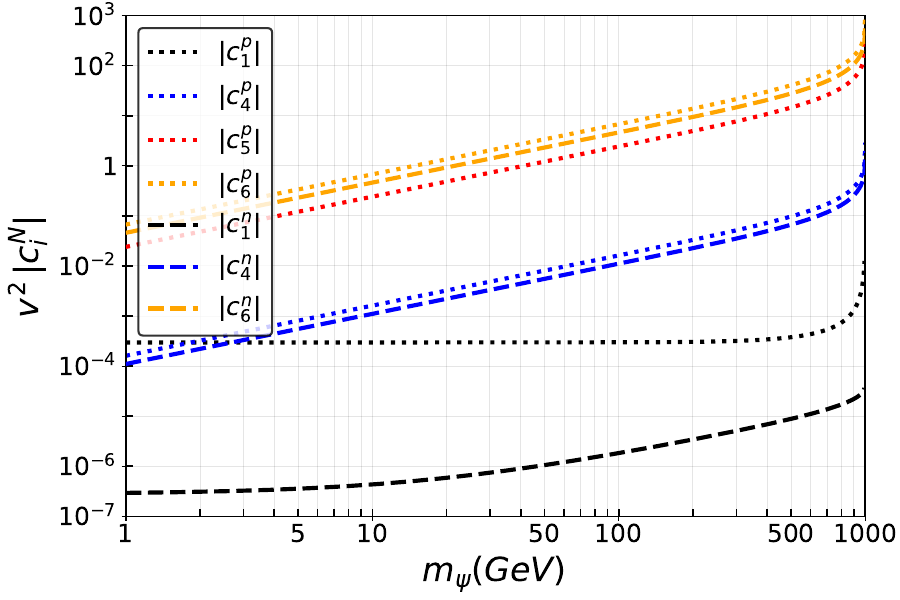}\hfill
		\includegraphics[width=0.48\linewidth]{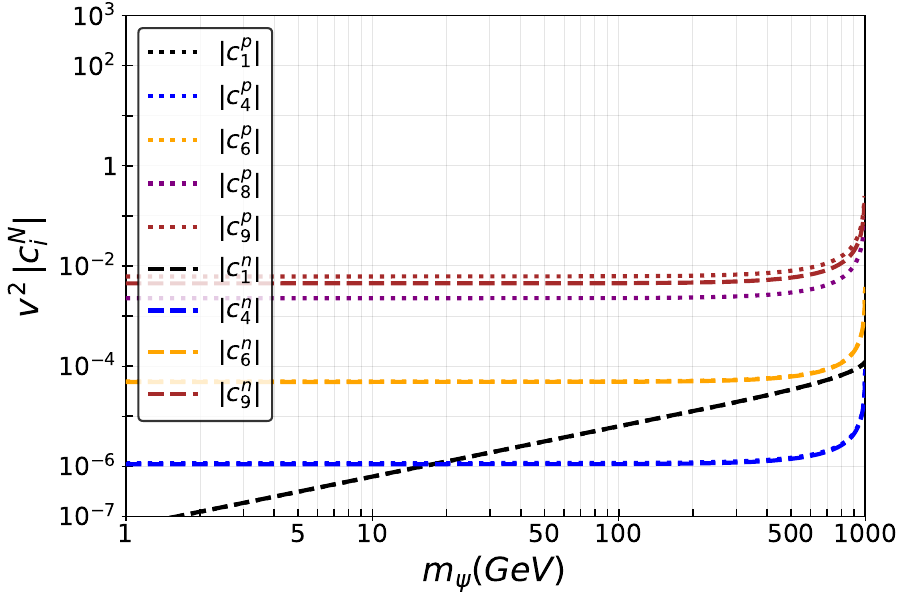}
		\caption{Dirac DM (left) and Majorana DM (right) with left-handed SM leptons in the loop and a scalar $S$  of mass $m_S=1000$ GeV. All Yukawa couplings are fixed to $y_2=1$.} 
	\end{subfigure}

	\caption{Non-relativistic nucleon level Wilson coefficients evaluated for ${}^{132}_{54}\mathrm{Xe}$ at $E_R=8.59$ keV (and thus $|\vec q|^2=2.11\times 10^{-3}\,\mathrm{GeV}$) versus the DM mass $m_\psi$.
	The Higgs portal coupling is $\lambda_{HS}=3$. All Wilson coefficients are displayed in dimensionless units by rescaling with the square of the electroweak VEV $v=246.2$ GeV.}
\label{fig:NR}
\end{figure}

The previous Wilson coefficients at the quark level can interfere and generate non-trivial effective operators 
 at the nucleon level, see Sec.~\ref{sec:WCnucleons}. We plot in Fig.~\ref{fig:NR} the NR Wilson coefficients with protons (neutrons) in dotted (dashed) lines ($N=n,\,p$ for neutrons and protons). All Wilson coefficients are displayed in dimensionless units, by rescaling them with the square of the electroweak VEV, $v=246.2$ GeV. As for Fig.~\ref{fig:cont}, the upper panel is for vector-like fermions and the lower panel for SM left-handed lepton doublets. The plots on the left correspond to Dirac DM, while the plots on the right are for Majorana DM. 

For a vector-like fermion $F$ (upper panels of Fig.~\ref{fig:NR}), we fix $m_F=600$
GeV, $m_S=500$ GeV, $y_V=1$ and $y_A=1.3\,e^{1.4\,i}$. For Dirac DM (left
plot), we show the coefficients short-range SI $c^N_1$ (black) and the SD scattering $c^N_4$ (blue), and the long-range contributions $c^N_5$ (red), $c^N_6$ (orange) and $c^N_{11}$ (green). Notice that both $c^N_5$ and $c^N_{11}$ are generated by the electric and magnetic DM dipole moments
proportionally to the nucleon charge and they are therefore absent for
neutrons. The long-range Wilson coefficients $c^N_5$, $c^N_6$ and $c^N_{11}$
dominate. The SD coefficients $c^N_4$ are more than two orders of magnitude smaller and very similar for protons and neutrons, although slightly larger for the former. The SI coefficients $c^N_1$ are the smallest ones, and $c^p_1$ decreases with the DM mass up to $m_\psi\simeq 500$ GeV. The difference in behavior of $c^p_1$ and $c^n_1$ stems mainly from the non-zero contribution
of $\mu_\psi$ to the former ($c^p_1 \propto \mu_\psi/m_\psi$). In this example the Wilson coefficients $c^N_1$ change sign at about $m_\psi\simeq 225$ GeV and $c^N_{4,5,6}$ at $m_\psi\simeq 425$ GeV.
For Majorana DM with vector-like fermions (top right) the  $c^N_1$ contributions generated by the Higgs penguin diagram (black) are very similar for protons and neutrons (they are superimposed in the plot). The anapole moment generates $c^p_8$ (solid purple), $c^p_9$ (solid magenta) and $c^n_9$ (dashed magenta) which are very similar, specially $c^p_8$, to the $c^N_1$ contributions (black). All the Wilson coefficients are in the range $10^{-5}-10^{-4}$, except in the region of DM mass when a Wilson coefficient changes sign at $m_\psi\simeq225$ GeV.

For Dirac DM with SM leptons (bottom left) the phenomenology
is very rich. The long-range Wilson coefficients $c^p_5$
(dotted red), $c^p_6$ (dotted orange) and $c^n_6$ (dashed orange) dominate
($c^n_5=0$, as it is proportional to the electric charge of the nucleon). They increase with the DM mass as they are generated dominantly
by $\mu_\psi$, i.e., chirality needs to be violated. Similarly $c^N_4$ (blue) with $p$ (dotted) and $n$ (dashed) have a dominant contribution from $\mu_\psi$ and therefore increase
with $m_\psi$. Regarding $c^n_1$ (dashed black), the
increase of its slope reflects the fact that the short-range Higgs contribution (which
grows with $m_\psi$) increasingly becomes more and more comparable to the
photon short-range coefficient, but in any case $c^n_1$ remains very
suppressed. $c^p_1$ is dominated by the photon penguin, and both the short-range contribution parameterized by $c_{VV,\gamma}^q$ and the long-range contribution from the magnetic moment $\mu_\psi/m_\psi$ are important. Due the dependence of the quark-level Wilson coefficients on the DM mass, the NR Wilson coefficient $c_1^p$ is basically constant with respect to it.
For Majorana DM with SM leptons (bottom right) the Wilson coefficients $c^p_8$ (solid purple), $c^p_9$ (solid magenta)
and $c^n_9$ (dashed magenta), which are generated by the anapole operator,
dominate. $c^N_6$  ($c^p_6$ and $c^n_6$ are superimposed in the plot) do not increase with the DM mass, unlike in the Dirac case, because here they come from $c^N_{\rm AA}$ and not from $\mu_\psi$; $c^N_1$, generated by the Higgs
penguin diagram, increases with $m_\psi$ and is very similar for $n$ and $p$ ($c^n_1$ and $c^p_1$ are superimposed in the plot). Finally, $c^N_4$, generated by the $Z$ penguin, are similar for both $n$ and $p$ (superimposed in the plot) and very suppressed, as expected.

\subsection{Direct detection event rates} \label{sec:bounds1}

\begin{figure}[tb]\centering
	\begin{subfigure}{0.48\linewidth}
		\includegraphics[width=\linewidth]{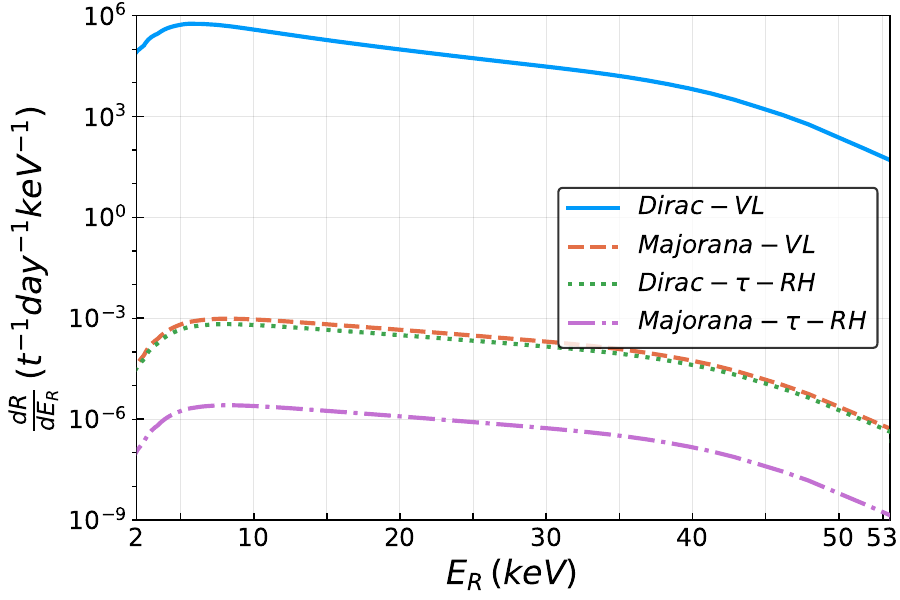}
		\caption{Dirac vs.~Majorana DM for a vector-like fermion and a right-handed $\tau$ lepton in the loop.} 
	\end{subfigure}
	\hfill
	\begin{subfigure}{0.48\linewidth}
	\includegraphics[width=\linewidth]{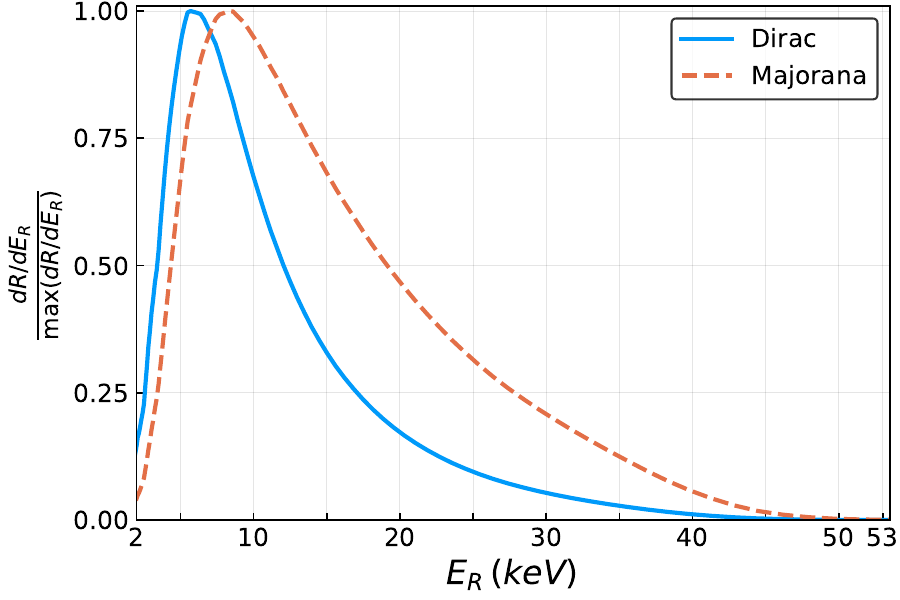}
	\caption{Normalized spectrum for Dirac and Majorana DM with a vector-like
		fermion in the loop.
	}
\end{subfigure}

\vspace{4ex}

	\begin{subfigure}{0.48\linewidth}
		\includegraphics[width=\linewidth]{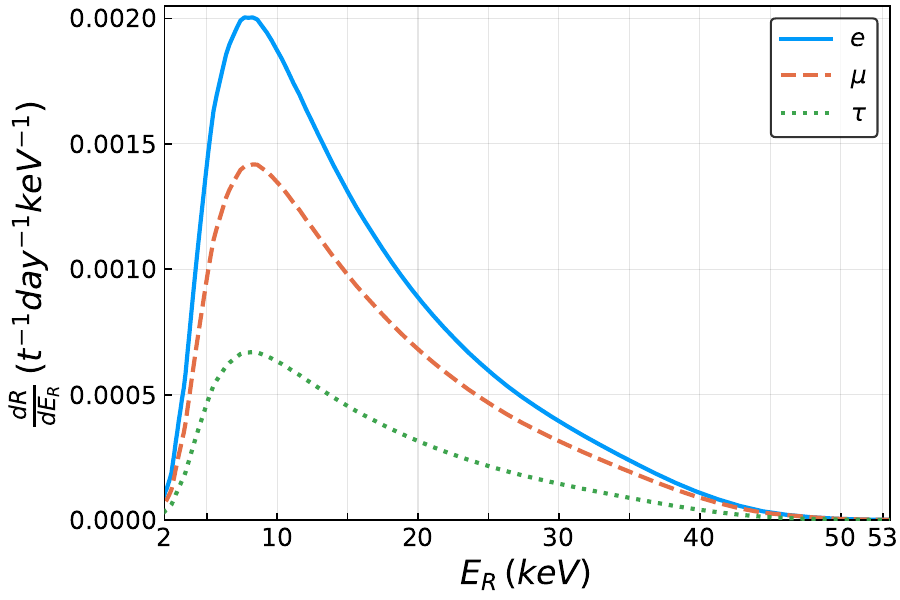}
		\caption{Dirac DM with coupling to a
		right-handed lepton. 
		A pure
	coupling to a left-handed lepton leads to the same result.}
	\end{subfigure}
	\hfill
\caption{Differential event rates for different combinations of DM candidates
and fermions in the loop. The DM mass is $m_\psi=90$ GeV and the Higgs portal coupling is $\lambda_{HS}=3$. In the
case of a  vector-like fermion $F$ and a scalar $S$ in the loop we fix
$m_F=600$ GeV, $m_S=500$ GeV, $y_V=1$ and $y_A=1.3\,e^{i\,1.4}$. In the case of
a right-handed $\tau$ lepton we fix $m_S=1000$ GeV and $y_1=1$.}
\label{Dirac_Maj}
\end{figure}

The different Wilson coefficients are expected to generate different features in the DD differential spectrum. In the upper-left panel of Fig.~\ref{Dirac_Maj}  we plot the DD differential
event rates in xenon versus the recoil energy $E_R$ for Dirac DM with a
vector-like fermion $F$ (solid blue) and with a right-handed tau (dotted green),
and for Majorana DM with a vector-like fermion (dashed red) and with a
right-handed tau lepton (dot-dashed purple). For details on the
astrophysical assumptions used in the numerical analysis see Ref.~\cite{Liu:2017kmx}. The rate for Dirac DM with a vector-like fermion is roughly $
9$ orders of magnitude larger than that with a SM lepton (a tau lepton in this case), because in the latter case there is no contribution from the electric dipole moment $d_\psi$ and the magnetic dipole moment $\mu_\psi$ is suppressed by
the DM mass $m_\psi$. The smallest rate occurs for Majorana DM with a
right-handed tau in the loop. The relative size of the spectra is obvious from the relative size of the NR Wilson coefficients discussed in the previous section.

In the upper-right panel we show the spectrum normalized to the maximum value ($5.7\times 10^{5}\,
[9.7\times 10^{-4}]\,{\rm t^{-1}\,day^{-1}\,{\rm keV}^{-1}}$ for Dirac [Majorana] DM)
 for the case with vector-like fermions in the loop, for Dirac DM (solid
blue) and Majorana DM (dashed red). The spectral shapes are
quite different, which is mainly due to the fact that there are no dipole moments for Majorana DM.

In the bottom panel of Fig.~\ref{Dirac_Maj} we plot the DD differential rates for Dirac DM with coupling to a right-handed electron (solid blue), muon (dashed red) and tau (dotted green). The spectra are the largest for the electron (the lightest lepton), with maxima at roughly the same recoil energy. The maxima go approximately in the ratios $\sim (4:2:1)$ for $e, \mu,\tau$. This is due to the dependence on the short-range contribution of the photon penguin, $c^q_{VV,\gamma}$, via the NR Wilson coefficient $c_1^p$. The spectra are dominated by the photon short-range contribution $c^q_{VV,\gamma}$ for this choice of parameters.

\subsection{Direct detection limits} \label{sec:bounds}

\begin{figure}[tb]\centering
	\begin{subfigure}{0.48\linewidth}
		\includegraphics[width=\linewidth]{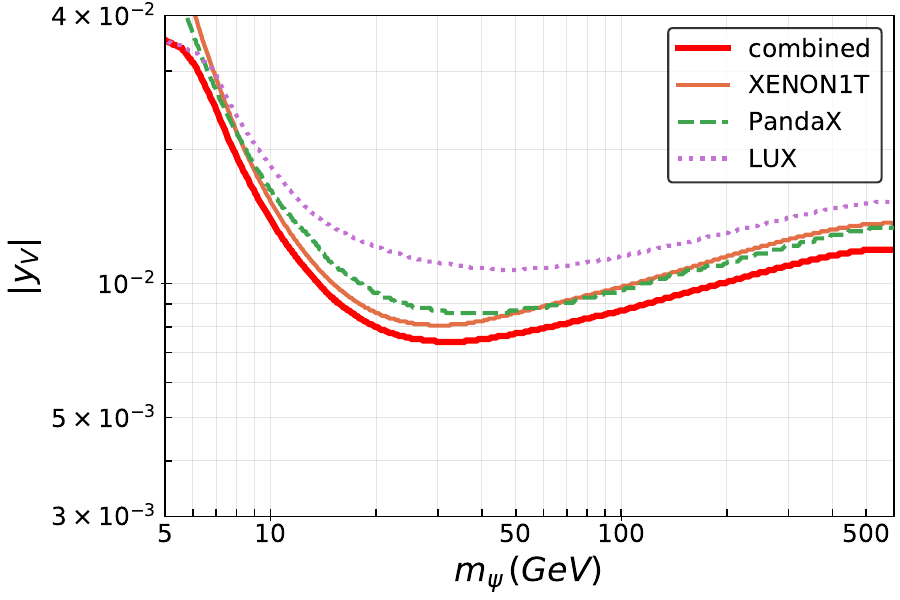}
		\caption{Dirac DM for different
		xenon experiments.}
	\end{subfigure}
	\hfill
	\begin{subfigure}{0.48\linewidth}
	\includegraphics[width=\linewidth]{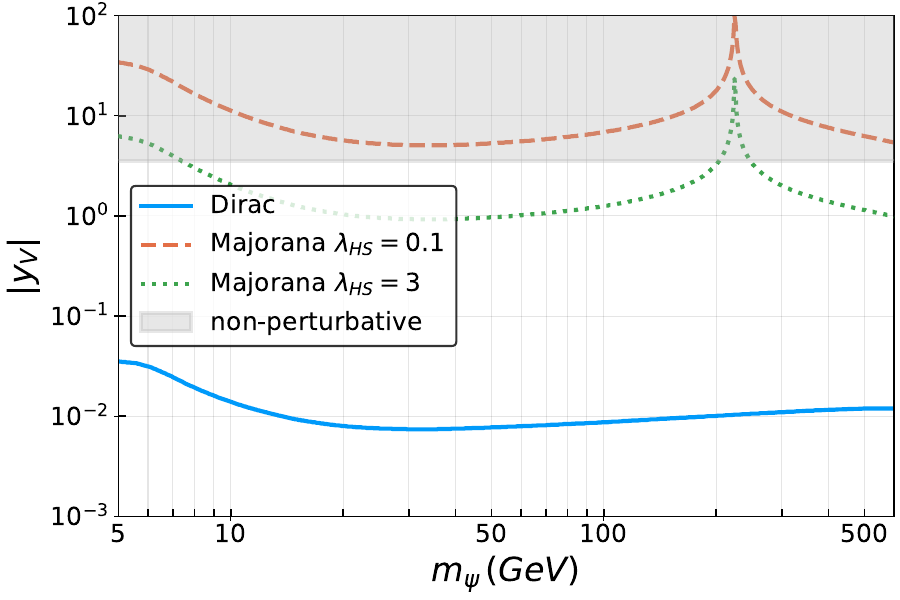}
	\caption{Different DM candidates and couplings. }
\end{subfigure}
	\caption{DM direct detection $90\%$ C.L. limits with a vector-like fermion in the loop. We fix $m_S=500$ GeV, $m_F=600$ GeV, the ratio of Yukawa
couplings $|y_A|/|y_V|=1.3$ and the phases of the Yukawa couplings are
$\phi_V=0$ and $\phi_A=1.4$.  Unless specified the Higgs portal coupling is
$\lambda_{HS}=0.1$. We highlight in gray the region where the Yukawa coupling is non-perturbative, $|y_V|>\sqrt{4\pi}$.}  \label{Dirac_Xe}
\end{figure}

Next we study the upper limits that current DD
experiments can impose on the scenarios discussed so far. In order to illustrate current direct detection limits, we consider different scenarios of TeV-scale dark sectors. We also discuss how the limits vary with the masses of the particles in the loop. We show the $90\%$ C.L. upper limits from current DD
experiments that have xenon as a target, which provide the
most stringent limits for SI interactions for our range of DM masses.  We show $m_\psi \gtrsim 5$ GeV, as very light DM does not produce recoils at energies above the threshold of the DD experiments. The
limits are subject to large uncertainties from nuclear physics and
astrophysics as well as from experimental uncertainties. 
In the following we do not show limits from Higgs and $Z$ boson invisible decay widths into DM, as those are weaker than the ones coming from DD in our scenarios. In Sec.~\ref{sec:ZHdisc} we discuss some examples where these limits can be relevant, and complementary to DD, specially for light DM masses, and in App.~\ref{ap:HZdecays} we provide the relevant expressions for the Higgs and $Z$ boson invisible decay rates.

In the left panel of Fig.~\ref{Dirac_Xe} we plot the upper limits for Dirac DM in the
plane $|y_V|$ versus $m_\psi$ for XENON1T (solid brown), PandaX (dashed green)
and LUX (dotted purple), together with their combined limit (thicker solid red
line). We have fixed $m_S=500$ GeV, $m_F=600$ GeV, $\lambda_{HS}=0.1$, the
ratio of Yukawa couplings $|y_A|/|y_V|=1.3$ and the phases of the Yukawa
couplings $\phi_V=0$ and $\phi_A=1.4$. As expected the bounds are weakened at
very large and very small DM masses. At large DM masses the limits appear to approach a constant value, instead of decreasing as $1/m_\psi$ as expected from the DM number density. This is due to the non-trivial dependence of the Wilson coefficients on $m_\psi$. In particular the Wilson coefficients generally increase for $m_\psi \rightarrow m_F+m_S$. The $|y_V|$ limits
are of the order of $\sim 10^{-2}$ for a large range of DM masses between 10 GeV
and 500 GeV. This is a clear example of the superb sensitivity achieved by DD
experiments, which are able to probe such small Yukawa couplings for
loop-induced scenarios of Dirac DM.

In the right panel of Fig.~\ref{Dirac_Xe} we show the limits for Majorana DM with $\lambda_{HS}=0.1$ (dashed red) and $\lambda_{HS}=3$ (dotted green), together with those for Dirac DM (solid blue). The current limits for Majorana DM are very weak, close to the naive perturbativity limit. Notice that the Higgs interactions are non-negligible:
changing $\lambda_{HS}=0.1$ to $\lambda_{HS}=3$ the upper bound on the Yukawa
couplings improves by a factor of $\sim 6$ (at the level of the rate, the
scalar quartic coupling enters quadratically, while the Yukawa couplings enter
to the fourth power). The difference with respect to the strong limits for
Dirac DM stems, of course, from the absence of dipole moments for Majorana DM.
In the gray shaded region, the Yukawa coupling is non-perturbative, $|y_V|>\sqrt{4\pi}$, and therefore the one-loop computation cannot be trusted.
 
\subsection{Interplay with lepton flavor violation and relic abundance} \label{sec:interplayLFV}
When there are SM charged leptons running in the loop, there may also be limits
from LFV processes. We provide the relevant expressions for $\ell_\alpha
\rightarrow \ell_\beta \gamma$, $\mu-e$ conversion and $\ell_\alpha \rightarrow
\ell_\beta \ell_\gamma \ell_\delta $ in App.~\ref{ap:LFV}.\footnote{In the following, we do not
show results for LFV Higgs and $Z$ boson decays, as the experimental limits on
these are weaker than limits from leptonic LFV decays.} It is therefore interesting to study
the interplay between both types of signals. Although one may naively expect that LFV limits are stronger (because an accidental symmetry of the SM is violated), we see in the following that this is not the case in all scenarios. 

\begin{figure}[p]
\centering  
		\begin{subfigure}{0.48\linewidth}
	\includegraphics[width=\linewidth]{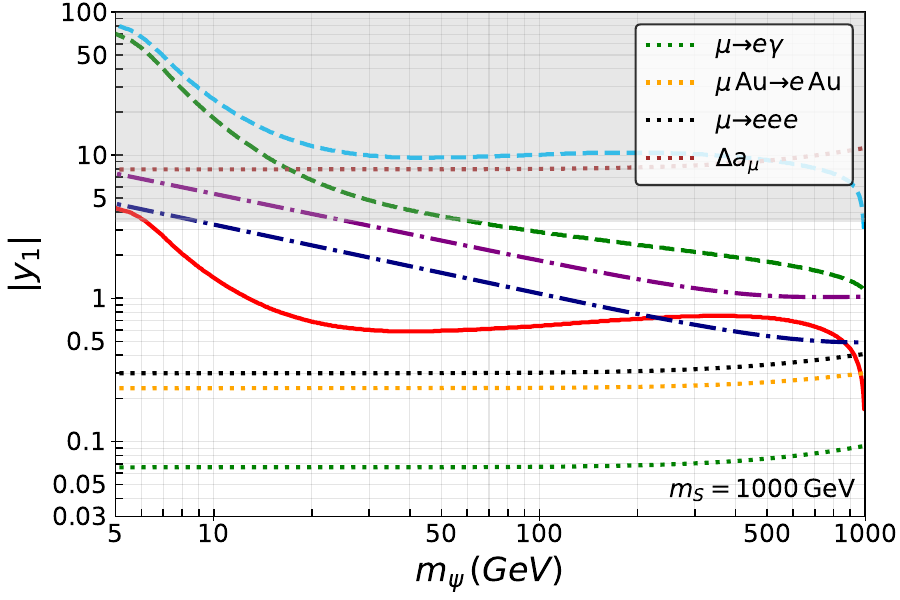}
	\caption{Equal leptons couplings. $m_S=1000$ GeV.}
\end{subfigure}
	\hfill
	\begin{subfigure}{0.48\linewidth}
	\includegraphics[width=\linewidth]{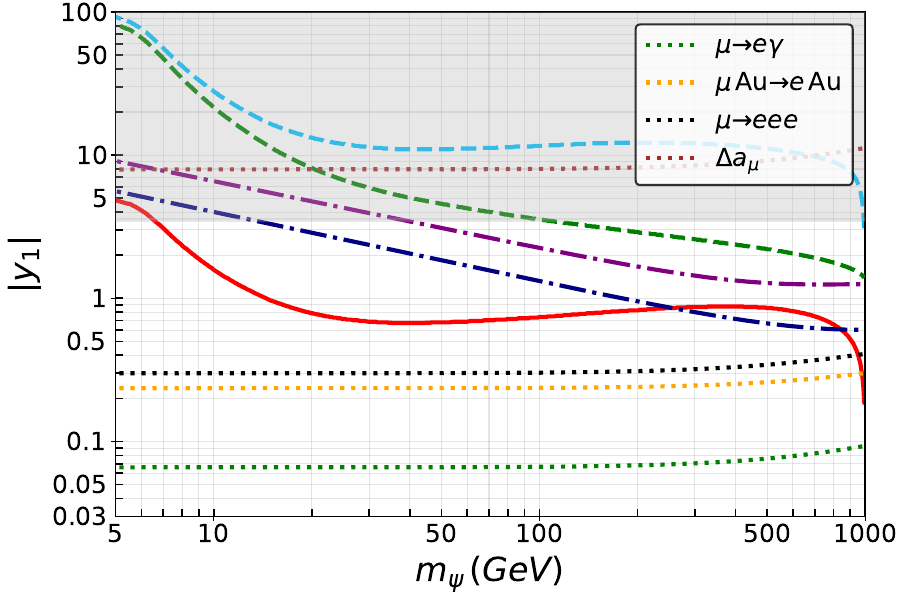}
	\caption{No coupling to $\tau$ leptons. $m_S=1000$ GeV.}
\end{subfigure}

\vspace{4ex}

	\begin{subfigure}{0.48\linewidth}
	\includegraphics[width=\linewidth]{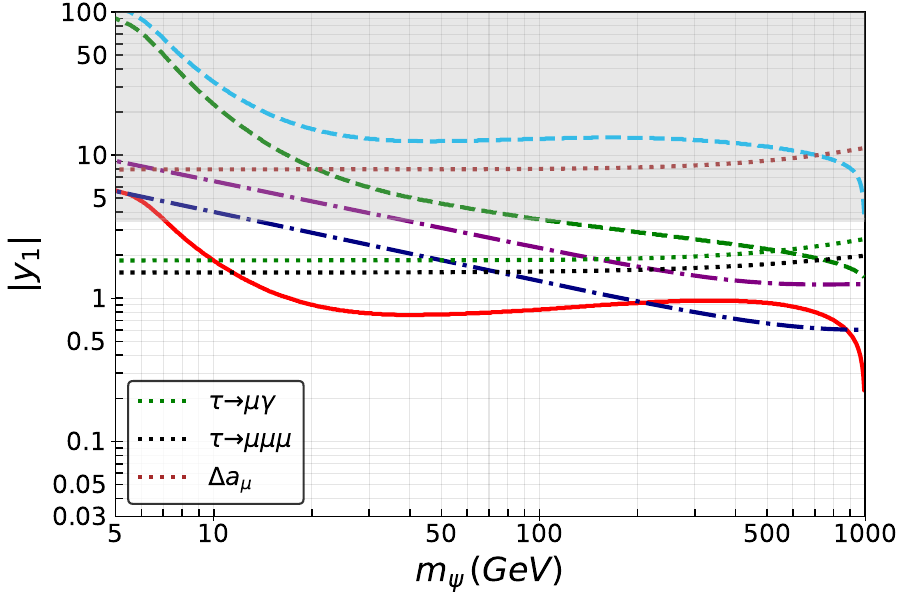}
	\caption{No coupling to electrons. $m_S=1000$ GeV.}
\end{subfigure}
\hfill
	\begin{subfigure}{0.48\linewidth}
	\includegraphics[width=\linewidth]{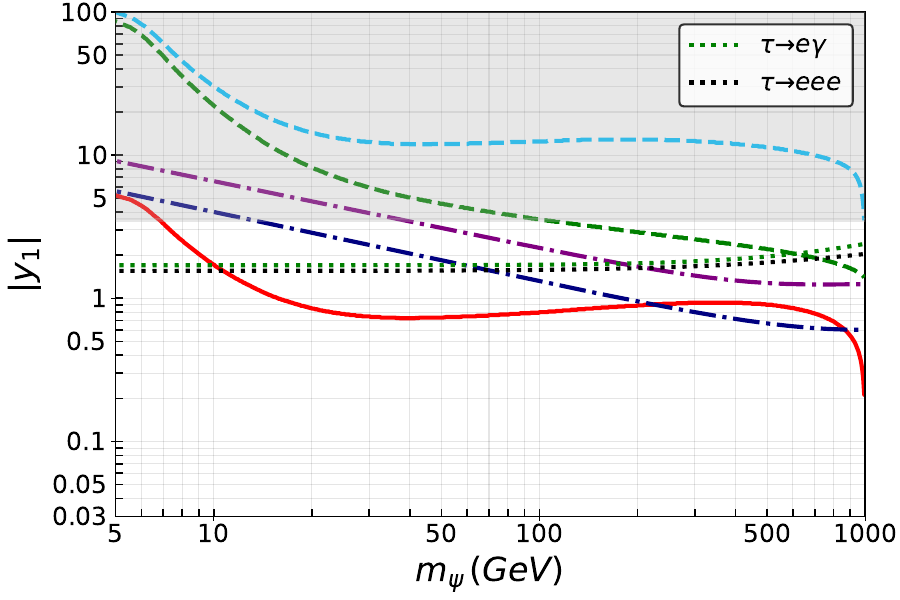}
\caption{No coupling to muons. $m_S=1000$ GeV.}
\end{subfigure}
	
\vspace{4ex}

\begin{subfigure}{0.48\linewidth}  
	\includegraphics[width=\linewidth]{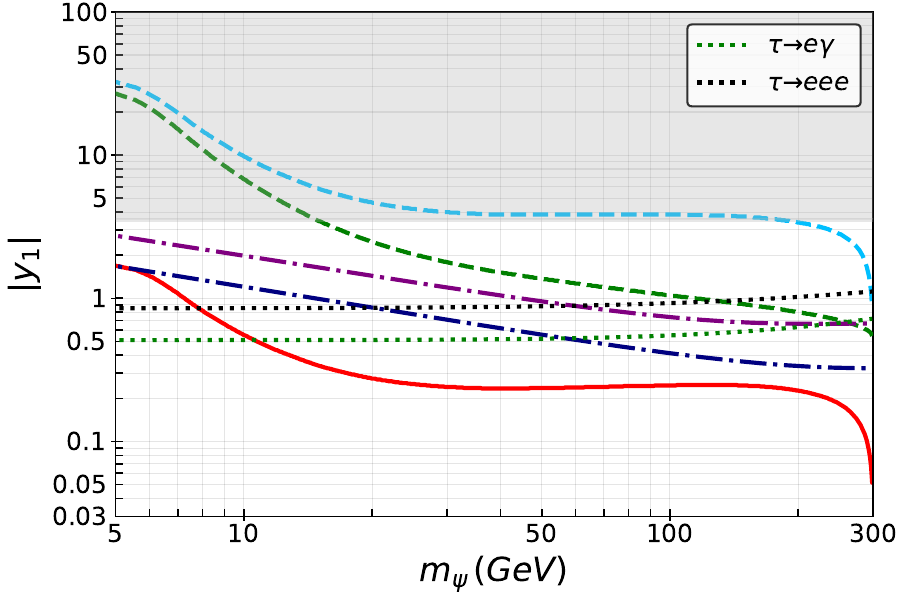}
\caption{No coupling to muons. $m_S=300$ GeV.}
\end{subfigure}
	\hfill
	\begin{subfigure}{0.48\linewidth}
	\includegraphics[width=\linewidth]{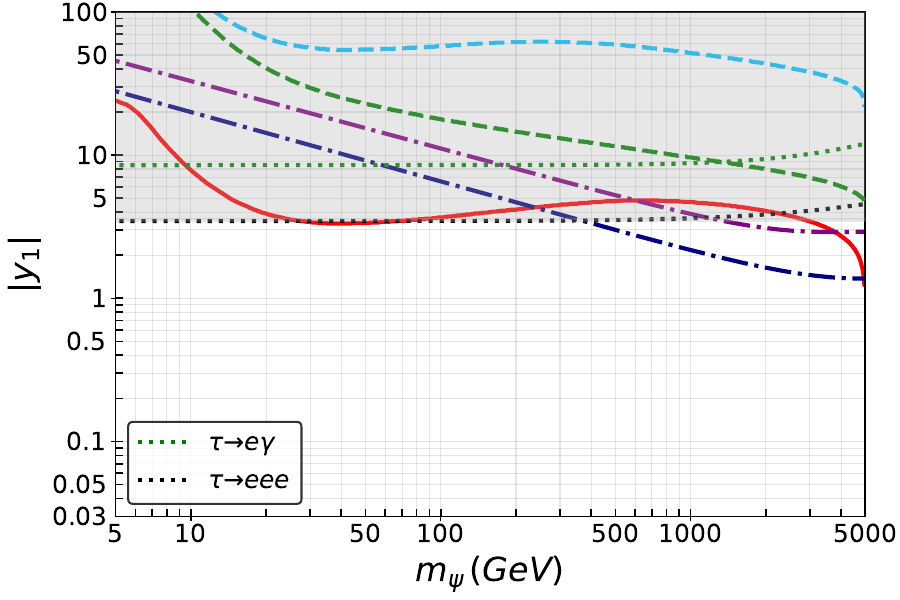}
\caption{No coupling to muons. $m_S=5000$ GeV.}
\end{subfigure}

\caption{Combined direct detection $90\%$ C.L. limits for Dirac dark matter (solid red), and Majorana dark matter with $\lambda_{HS}=0.1$ (dashed light blue) and $\lambda_{HS}=3$ (dashed green), with right-handed charged leptons in the loop. Contours of the correct relic abundance set by $\psi \psi \rightarrow \ell_\alpha\ell_\beta$ annihilations mediated by the scalar $S$ are shown as dot-dashed navy blue (purple) line for Dirac (Majorana) dark matter. The dotted lines indicate constraints from the relevant LFV processes. In the gray shaded region the Yukawa coupling is non-perturbative, $|y_1|>\sqrt{4\pi}$. } \label{equal_notau}
\end{figure}

In Fig.~\ref{equal_notau}, top-left panel, we plot the DD upper limits\footnote{In the following we only show the combined limit from all xenon
experiments, like the thicker solid red line shown in the left panel in
Fig.~\ref{Dirac_Xe}, but for the case of SM leptons in the loop.} in the plane
$|y_1|$ versus $m_\psi$, assuming equal couplings to all leptons, i.e.,
$y^e_1=y^\mu_1=y^\tau_1=y_1$  (we denote this the ``symmetric" case). In Fig.~\ref{equal_notau} top-right, middle-left and middle-right panels we
show the cases of no couplings to taus, electrons and muons, respectively. Left-handed and right-handed leptons in the loop lead to the same result. We show
the cases of Dirac DM (solid red), and Majorana DM with $\lambda_{HS}=0.1$
(dashed light blue) and $\lambda_{HS}=3$ (dashed green). We have fixed $m_S=1000$ GeV
for the four upper plots. The most relevant $90\%$ C.L. LFV limits are shown
using dotted lines: $\mu \rightarrow e \gamma$ (green), $\mu-e$ conversion (orange), $\mu \rightarrow 3e$
(black) and $\Delta a_\mu$ (brown).\footnote{This corresponds to the $4\sigma$ limit
coming from the AMM of the muon $\Delta a_\mu$. This discrepancy with respect
to the SM cannot be explained in our model, because the additional contribution
is negative and thus leads to a larger departure from the experimental value.} Notice that LFV limits do not depend
on whether the DM is a Dirac or Majorana fermion. Also, we emphasize once more that DD limits are subject to large astrophysical and nuclear uncertainties, which are absent in the case of LFV experiments.

In addition we plot the contour of the DM relic abundance, set by t-channel DM annihilations $\psi \psi \rightarrow \ell_\alpha\ell_\beta$ mediated by the scalar $S$, with a dot-dashed navy blue (purple) line for Dirac (Majorana) DM, whose leading contribution is from s-wave (p-wave) scattering. We use the instantaneous freeze-out approximation which is sufficient for our purposes (see Sec.~\ref{sec:relicth} and App.~\ref{ap:relic_details} for more details and the relevant
expressions). Above the $\Omega h^2$ contour the DM would be under abundant and requires an additional
component of DM to account for the observed relic abundance. Below the $\Omega
h^2$ contour DM is over abundant if its abundance is solely set by freeze-out,
and thus there has to be a mechanism to further deplete its density.
It could be reduced via co-annihilation and resonant effects
\cite{Griest:1990kh}, multi-body scatterings \cite{Dolgov:1980uu, Dey:2016qgf,
Choi:2017mkk,Cline:2017tka, Bell:2017irk}, or a non-trivial thermal evolution
in the early universe~\cite{Kobakhidze:2017ini}. In case $\psi$ does not account for all of the DM
abundance the DD limits have to be rescaled appropriately. Assuming thermal freeze-out reproducing the correct relic abundance imposes a lower
bound on the DM mass. In the case of equal couplings to all
leptons with $m_S=1000$ GeV, $m_\psi \gtrsim 10\,(25)$
GeV for Dirac (Majorana) DM. When one final channel is
closed, the lower limits increase by roughly $5\,(15)$
GeV for Dirac (Majorana) DM. For light scalar mass (see
bottom-left panel of Fig.~\ref{equal_notau}), all
Yukawa couplings are perturbative. However, for heavy $m_S$,
bottom-right panel, the Yukawa couplings are perturbative only for
very heavy masses, above $0.4\,(1)$ TeV, as in this case the
t-channel interaction is significantly suppressed by the mass of
the mediator. 

The main changes in the case of no couplings to taus, electrons and muons (top-right, middle-left and middle-right panels in Fig.~\ref{equal_notau}) are in the
LFV limits, as depending on the flavor structure, different processes are
possible. In these panels the relic abundance contours are almost identical, as
the SM leptons are always much lighter than the DM (and therefore phase space
plays no significant role). Of course, the contours are at somewhat larger
Yukawa couplings than for the ``symmetric'' scenario, as in the latter there
were more available annihilation channels. The DD limits are also
slightly modified due to the different masses of SM leptons in the loop (see
also the lower panel of Fig.~\ref{Dirac_Maj}). When there are no couplings to
taus (top-right panel), the LFV limits are almost identical to the ``symmetric''
scenario, because they are driven by the first family. However, for no couplings to electrons or muons (middle panels), DD limits are more stringent than LFV limits for Dirac DM with a mass above a certain value. 
This is quite remarkable: DD experiments are able to better constrain scenarios where an accidental symmetry of the SM is violated than experiments directly searching for it.
Interestingly, limits on $|y_1|$ from trilepton $\tau$ decays ($\tau
\rightarrow 3 \ell$) dominate over radiative $\tau$ decays ($\tau\to \ell
\gamma$) in contrast to the limits from muon decays. As the limits from $\tau$
decays are generally weaker and thus the corresponding Yukawa couplings larger,
box-diagram contributions to trilepton decays may give a sizable contribution
and thus break the dipole dominance.

A few interesting remarks can be drawn from these plots. First, note that the
DD limits with SM leptons in the loop, even for Dirac DM, are much weaker than
in the scenario with vector-like fermions in the loop, as also demonstrated in
the top-left panel of Fig.~\ref{Dirac_Maj}. Second, clearly the LFV limits are
the strongest ones, with $\mu \rightarrow e \gamma$ the most stringent among them. Its limit on the Yukawa
coupling $|y_1|$ is a factor of a few stronger than the one of DD for Dirac DM. Again, the DD limits become very
strong close to $m_\psi\rightarrow m_S+m_F$ as in the case with vector-like
fermions. Third, for scalar masses at the TeV scale, the DD limit already excludes the production via thermal freeze-out for Majorana DM, and also for Dirac DM in the mass range $5~\text{GeV} \lesssim m_\psi \lesssim 200$ GeV.  Finally, the muon AMM constraint is always very weak, being the limit above the perturbativity bound. 

In Fig.~\ref{equal_notau}, bottom panels, we show two examples of a scalar $S$ in the loop with a different mass: $m_S=300$ GeV (left plot) and $m_S=5000$ GeV (right
plot). All limits are generically stronger for $m_S=300$ GeV and weaker for
$m_S=5000$ GeV compared to $m_S=1000$ GeV. In particular, the relative
contribution of the box diagrams and the dipole moment for the trilepton $\tau$
decay changes: for $m_S=300$ GeV $\tau \rightarrow e\gamma$ sets a
stronger limit than $\tau\to 3e$. Similarly the Yukawa coupling required to
explain the observed relic abundance also has to be larger for heavier scalar
masses, as already discussed. Indeed, for  $m_S=5000$ GeV almost all the limits on the Yukawa couplings are in the non-perturbative region.

In summary, strong limits can be set for Dirac DM with vector-like fermions in the loop. For Dirac DM with SM leptons in the loop LFV
limits or DD limits may set the strongest bounds depending on the flavor
structure and the DM mass. Therefore, the two limits are complementary: 
LFV limits are more important for DM coupling to both muons and electrons, whereas DD limits dominate if there are no LFV processes of type $\mu\to e X$, $X$
being anything, and the DM mass is not too small ($m_\psi\gtrsim 5$ GeV). For Majorana DM, LFV limits, if present, are generally more stringent than constraints from DD. Future DD experiments and LFV limits on $\tau$ decays are expected to improve by 1-2 orders of magnitude and hence the situation is not expected to change dramatically. 
If $\mu-e$ conversion in nuclei and/or $\mu \rightarrow 3e$
expected sensitivities (by several orders of magnitude) are achieved, LFV
limits will continue to dominate and even increase their difference with
respect to DD.

\section{Other phenomenological aspects} \label{sec:other}

\subsection{LHC searches}
Generally colliders may only set competitive limits via missing energy searches for light DM and SD interactions. In the scenarios discussed here, naively the production of DM particles at the LHC occurs at one-loop level via the penguin diagrams in Fig.~\ref{fig:penguins} and is therefore suppressed. For example, Ref.~\cite{Dissauer:2012xa} showed that there are only very weak collider limits on a model with a magnetic moment interaction. Thus it is more promising to search for the mediators $S$ and $F$ at colliders via $q  \bar q \rightarrow F  \bar F, S S^*$ mediated by the photon, the $Z$ boson and/or the Higgs. If the new fermion and scalar have electric charge, the production is dominated by the Drell-Yan process. Higgs-mediated production of exotic particles has been discussed in e.g.~Ref.~\cite{Hessler:2014ssa}.  As we are assuming that the new particles are not colored, only modest lower limits (below $1$ TeV) are expected, unless very large SM quantum numbers (for instance electric charges) are invoked. The dark sector particles may decay invisibly into DM and a lighter dark sector state. The phenomenology of these decays are however model-dependent, see discussion in Sec.~\ref{sec:setup}. Another interesting option would be to search for DM in models with electrons/muons running in the loop at future lepton colliders. The main production process is via $t$-channel exchange of the scalar, $\ell^+  \ell^- \rightarrow \psi  \bar \psi$ with $\ell=e,\mu$.

\mathversion{bold}
\subsection{$Z$ and Higgs boson invisible decays} \label{sec:ZHdisc}
\mathversion{normal}

\begin{figure} [t!] \centering
	\begin{subfigure}{0.47\linewidth}
		\includegraphics[width=\linewidth]{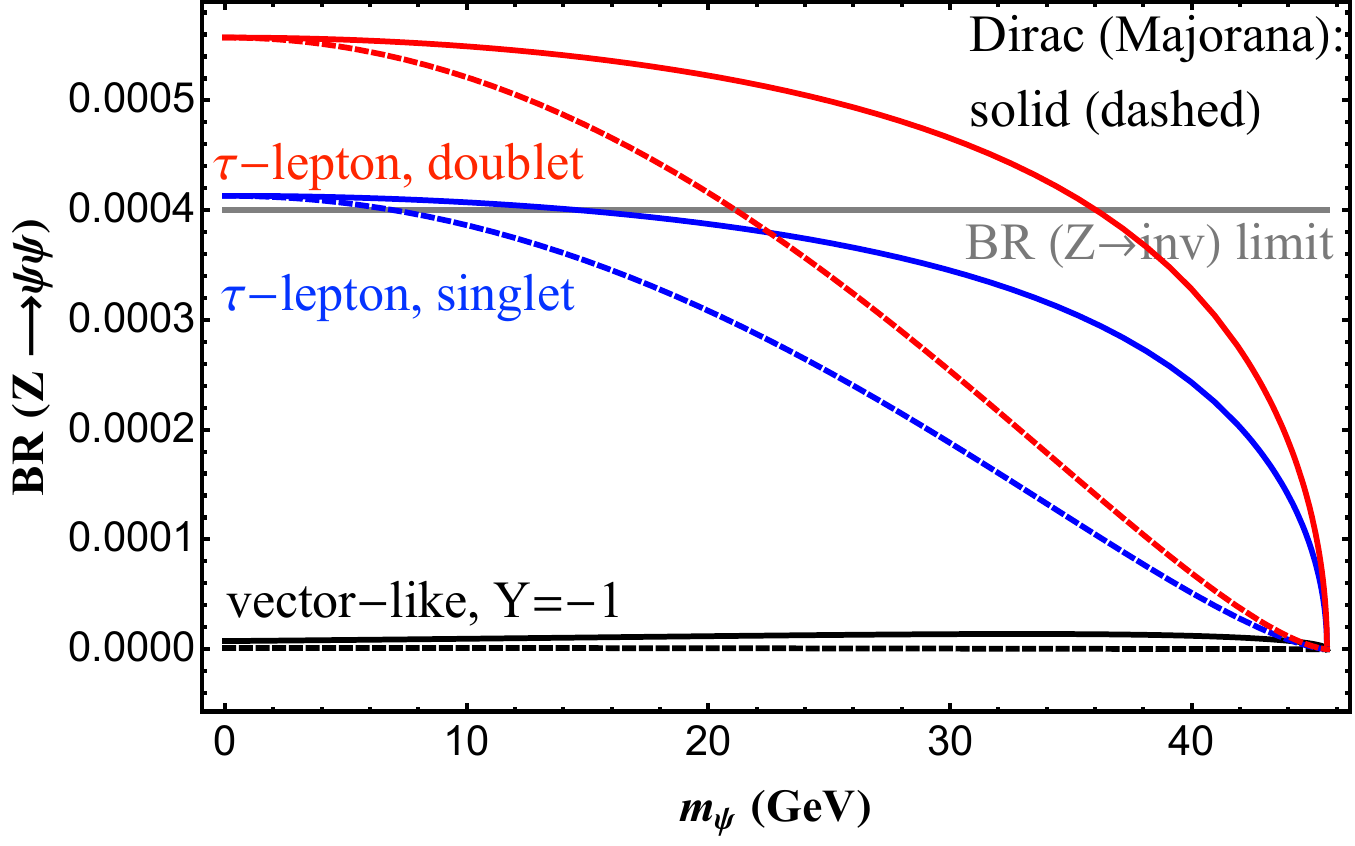}
		\caption{Branching ratio of the $Z$ boson into DM.}
	\end{subfigure}
	\hfill
	\begin{subfigure}{0.44\linewidth}
	\includegraphics[width=\linewidth]{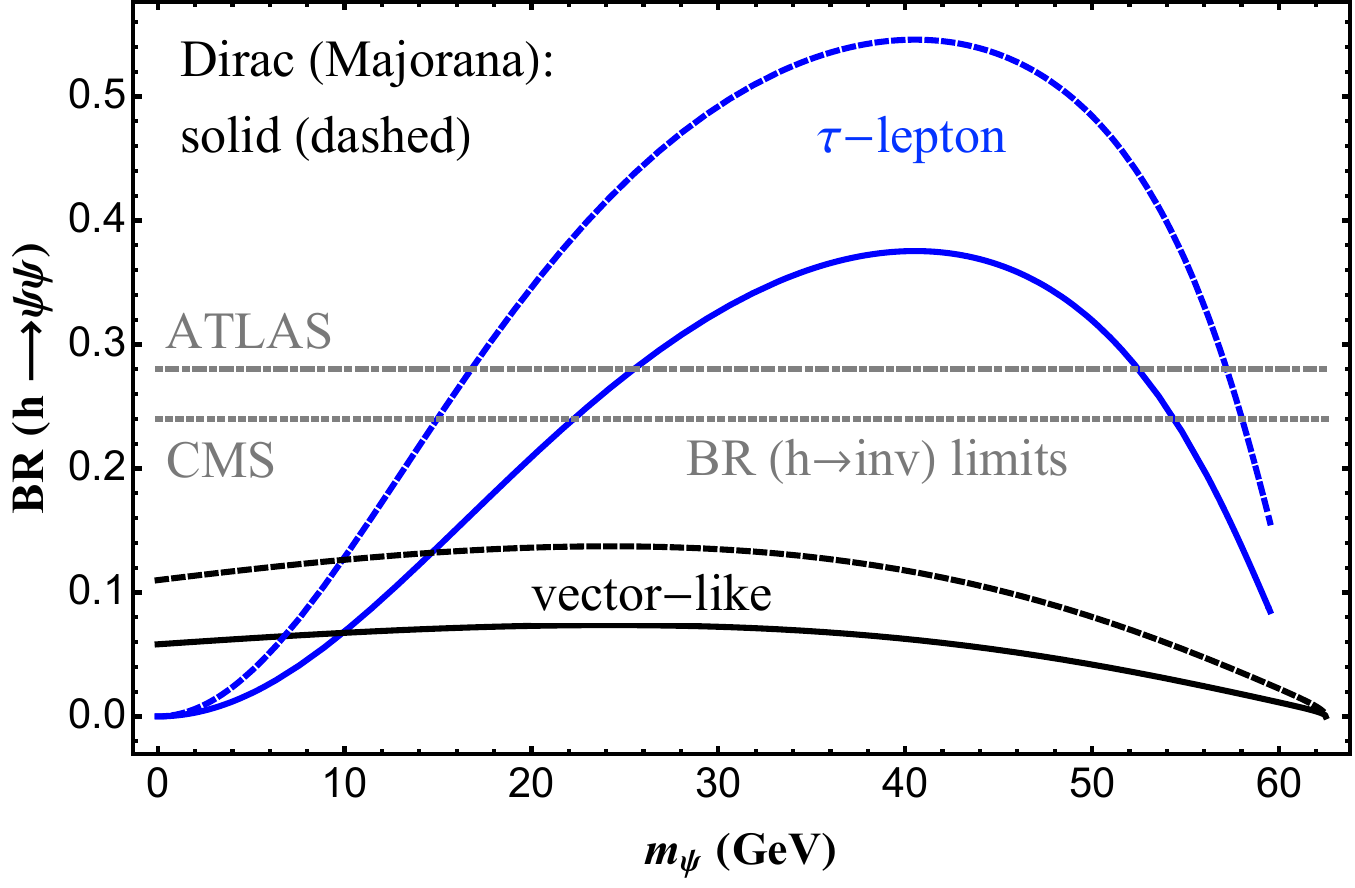}
	\caption{Branching ratio of the Higgs boson into DM.}
\end{subfigure}
	\caption{Branching ratios of the $Z$ and the Higgs bosons decaying invisibly into DM (Dirac in solid, Majorana in dashed). We show in black the case of vector-like fermions in the loop, in red the case of a tau-lepton doublet and in blue the case of tau-lepton singlet. The experimental upper limits on non-SM invisible decays are displayed as horizontal gray lines. See the text for details.}  \label{Decays_inv}
\end{figure}

If the DM $\psi$ is sufficiently light [$m_\psi<m_H/2\,\left(m_Z/2\right)$] there is an additional contribution to the invisible width of the Higgs ($Z$) boson. In App.~\ref{ap:HZdecays} we present the relevant expressions for these processes. We find that there are no limits from $Z$ or Higgs boson decays into DM for the parameter values used in Figs.~\ref{Dirac_Xe} and \ref{equal_notau}. However, there may be limits for small scalar/fermion/DM masses and large Yukawa couplings. To illustrate this point we plot in Fig.~\ref{Decays_inv} the branching ratios ${\rm Br}(Z \rightarrow \psi \psi)$ (left plot) and ${\rm Br}(h \rightarrow \psi
\psi)$ (right plot), for Dirac (Majorana) DM with solid (dashed) lines. For the SM widths we use $\Gamma_{h, {\rm SM}}=4.1$ MeV and
$\Gamma_{Z, {\rm SM}}=2.495$ GeV, such that the Higgs branching ratio reads ${\rm
Br}(h \rightarrow \psi \psi)=\Gamma_{h \rightarrow \psi \psi}/(\Gamma_{
h, SM}+\Gamma_{h \rightarrow \psi \psi})$ and similarly for the $Z$ boson. 
We show the cases of different particles running in the loop with solid lines:
in black the case of vector-like fermions with $Q_F=Y_F=-1$ and in red (blue)
the case of a tau-lepton doublet (singlet). For Higgs decays the tau-lepton
doublet and the singlet generate the same branching ratio, shown in blue. The experimental upper limits on invisible non-SM decays are shown as horizontal gray lines: solid for the $Z$ boson from LEP (the total invisible width of the $Z$ including neutrinos is $\Gamma_{\rm Z \rightarrow {\rm inv}}= 499.1 \pm1.5$ MeV~\cite{Patrignani:2016xqp}), and dot-dashed (dashed) for the Higgs from CMS \cite{Khachatryan:2016whc}  (ATLAS \cite{Aad:2015txa}), which reads ${\rm Br}(h \rightarrow {\rm inv})<0.24\,(0.28)$ at $95\%$ CL. We used $m_S=120$ GeV and $m_F=150$ GeV and a Higgs portal coupling $\lambda_{HS}=0.2$. For vector-like fermions in the loop we used $y_V=4\,e^{i\pi/3}$ and
$y_A=3\,e^{i\pi/4}$, while for SM tau-lepton doublets [singlets] we fixed
$y_V=[-]\,y_A=4\,e^{i\pi/3}$. 

In Fig.~\ref{Decays_inv} one can observe that the limits for Dirac DM are stronger than those for Majorana DM in the case of $Z$ boson decays, independently of the particles in the loop, while the situation is the opposite in the case of Higgs decays. Also, invisible $Z$ boson decays constrain light DM which couples to SM leptons (the tau in this case). For Dirac DM the limits exclude DM masses below $14\,(36)$ GeV in the case of couplings to tau singlets (doublets). The width is dominated by $c_V$ and $c_A$, while $d_A\simeq0$ and $d_V$ is suppressed by
$m_\psi$. For vector-like fermions the width is dominated by $d_V$ and $d_A$ with $d_V>d_A$, and there are no relevant limits. For Majorana DM the limits are weaker than for Dirac DM, demanding $m_\psi \gtrsim 6\,(21)$ GeV in the case of couplings to tau singlets (doublets), with no limits in the case of couplings to vector-like leptons. 

As in the case of the $Z$ boson, the decays of the Higgs boson do not pose limits on the scenario with vector-like fermions
in the loop. For the tau-lepton $b_A=0$ and the dominant contribution to $b_V$
is proportional to $m_\psi$, as $m_F \ll m_\psi$. The branching ratio
increases with the DM mass for low DM masses, while at some DM mass value ($\simeq 40$ GeV in
the plot) the phase space suppression dominates and the branching ratio decreases again.
Therefore there is a constraint on an intermediate DM mass range of $[25,\,53]$ GeV ($[22,\,55]$ GeV) by ATLAS (CMS) for Dirac DM and $[16,\,57]$ GeV ($[14,\,58]$ GeV) by ATLAS (CMS) for Majorana DM.

To summarize, while for vector-like fermions there are no limits, for SM particles in
the loop there may be interesting constraints in the absence of LFV. Indeed, there is a well-known complementarity between invisible decays and DD. The experimental energy threshold of DD experiments limits their ability to impose limits for arbitrarily low DM masses and thus invisible decays may set competitive limits for low DM masses.

\subsection{Relic abundance} \label{sec:relic}

The production of the correct relic DM density in the early universe is generally model-dependent. Although it is not the main focus of this work, we briefly outline different avenues to obtain the correct relic density.
See  e.g.~Ref.~\cite{Blennow:2015gta} for a connection of DD with thermal freeze-out.

\subsubsection{Thermal freeze-out} \label{sec:relicth}
If $m_\psi >m_S, m_F$ (but of course $m_\psi <m_S + m_F$), the relic abundance can be set via the t-channel interactions
$\psi \bar \psi \rightarrow S S^*$ or $\psi \bar \psi \rightarrow F \bar F$.
Subsequently, $S$ and $F$ can decay to SM particles, in some cases at loop
level or via non-renormalizable operators. In particular if $F$ is a SM
lepton $\ell_\alpha$, DM annihilations to SM leptons $\psi \bar \psi \rightarrow \ell_\alpha \bar
\ell_\beta$ may set the relic abundance. For Dirac DM the cross section is not velocity suppressed and thus the leading (s-wave) part of the thermally averaged
annihilation cross section\footnote{
The thermally averaged cross section $\braket{\sigma v}= a + 6 b/x$ with $x=m_\psi/T$ is obtained by integrating over the annihilation cross section 
$\sigma v= a+ b v^2$, after it has been expanded up to second order in the relative of velocity of the two DM particles in the center of mass frame $v=|\vec v|$.
Note that, although the DM is
non-relativistic at freeze-out, the relative velocity is not small,
$v_f=\sqrt{12/x_f}\simeq 0.7\,c$ in terms of the speed of light $c$.} is given by
\begin{equation} \label{eq:Dann}
	\left.\braket{\sigma v}\right|_\text{D}= \frac{m_\psi^2}{32\pi (m_\psi^2+m_S^2)^2}\, \sum_{\alpha,\beta} |y_{i,\beta} y_{i,\alpha}^* |^2\,,
\end{equation}
where we have summed over all possible final state leptons (neutrinos and
charged leptons) in the limit of vanishing lepton masses. Here $i=1\,(2)$ for
couplings to LH (RH) leptons, see Eq.~\eqref{Lpsid}. For
Majorana DM the annihilation cross section is velocity suppressed and the leading contribution is due to p-wave scattering\footnote{
	Annihilation channels with 3-body final states which lift the velocity suppression are generally not important during freeze-out due to the additional phase space suppression, but they are very important for indirect detection. Their importance for indirect detection has been pointed out in several papers~\cite{Bergstrom:2008gr,Bell:2011if}, see also
Refs.~\cite{Bringmann:2012vr,Kopp:2014tsa}.}
\begin{equation}  \label{eq:Mann}
	\left.\braket{\sigma v}\right|_\text{M} = \frac{m_{\psi}^2 \left(m_S^4+m_{\psi }^4\right)}{8 \pi x \left(m_S^2+m_{\psi}^2\right)^4}\,  \sum_{\alpha,\beta} |y_{i,\beta} y_{i,\alpha}^* |^2\,,\\
\end{equation}
where $x=m_\psi/T$. 

As discussed in App.~\ref{ap:relic_details}, for DM masses in the range $10\,\text{GeV}\lesssim m_\psi \lesssim 10^4\,\text{GeV}$ we obtain the correct relic abundance for cross sections $\left.\braket{\sigma v}\right|_\text{D} \simeq [2, 3] \cdot 10^{-26}\, {\rm
		cm^3\, s^{-1}}$ for Dirac DM and $\left.\braket{\sigma
		v}\right|_\text{M} \simeq [0.5, 1] \cdot 10^{-23}\, {\rm cm^3\,
	s^{-1}}$ for Majorana DM. Equating these values to Eq.~\eqref{eq:Dann}
	and Eq.~\eqref{eq:Mann}, respectively, we plot in Fig.~\ref{equal_notau} the relic abundance contours in the $|y_1|-m_\psi$ plane. 
	
If $\psi$ is the lightest particle in the dark sector (i.e.,
$m_\psi <m_S, m_F$), DM may annihilate at one-loop order into quarks via the penguin diagrams in Fig.~\ref{fig:penguins}. However this is very suppressed and results in an over abundance of DM and requires another mechanism:
(i) In a larger dark sector DM may annihilate into other lighter dark particles, $\psi \psi \rightarrow X X$ which subsequently decay to SM particles. These new light particles may lead to large DM self-interactions, see for instance Ref.~\cite{Blennow:2016gde}.
(ii) Co-annihilation and resonant effects~\cite{Griest:1990kh} may increase the effective thermal annihilation cross section. For example processes like $\psi \bar F \rightarrow S^* \rightarrow H H$ with $(m_F-m_\psi)/m_\psi\simeq 1/20$ could be induced by a coupling of $S$ to the SM Higgs.\footnote{If $S$ carries a dark charge it may be a soft-breaking term.} Similarly there may be coannihilations with $S$. If $S$ has gauge interactions the dominant channel may be $SS \to \text{SM\,SM}$~(see for instance Ref.~\cite{Khoze:2017ixx}) if $(m_F-m_\psi)/m_\psi\simeq 1/20$ and $\psi$ and $S$ are in thermal equilibrium.
(iii) Multi-body scatterings may also increase the effective thermal annihilation cross section~\cite{Dolgov:1980uu, Dey:2016qgf,
Choi:2017mkk,Cline:2017tka, Bell:2017irk}.
(iv) A non-trivial thermal evolution in the early universe may depopulate an initially over abundant DM relic density~\cite{Kobakhidze:2017ini}.

\subsubsection{Non-thermal production}
The DM abundance may also be produced non-thermally. If DM is only very weakly coupled to the SM thermal bath and it has not been produced during reheating, DM may be slowly produced via the freeze-in mechanism~\cite{McDonald:2001vt,Hall:2009bx}. Ref.~\cite{Molinaro:2014lfa} discussed the phenomenology of the freeze-in mechanism in the scotogenic model~\cite{Ma:2006km} with fermionic DM, one of the examples where DM-nucleus scattering occurs at one-loop level.

\section{Conclusions} \label{sec:conc}

Direct detection of DM may not have been observed yet because it is absent at tree level, occurring only at the loop level. In this work we have studied the case of a fermionic singlet DM $\psi$, which is a
simple scenario where DD is naturally induced at one-loop order. The type of scenario considered
appears in supersymmetric extensions where the neutralino is
pure bino~\cite{Berlin:2015njh} (notice that in this case its mass is
typically very heavy, larger than 2 TeV), and also in connection to
neutrino masses, in particular in the seesaw model~\cite{Gonzalez-Macias:2016vxy} and in some radiative neutrino mass
models~\cite{Ma:2006km,Schmidt:2012yg, Ibarra:2016dlb, Hagedorn:2018spx}. We have
considered a simplified scenario with a dark sector made of a
vector-like (or a SM) fermion and a (complex) scalar. We presented
general analytical expressions for the different contributions as well
as current limits on the dark sector parameters.
We have outlined the possible UV completions of the corresponding
penguin diagrams, also those involving SM fields, and we summarize the different possibilities in the following:

(i) If the fermion is a SM lepton and thus leptophilic, the DM interactions are generically flavored~\cite{Agrawal:2011ze} and there is an interesting phenomenology. 
There may be new contributions to the anomalous magnetic moment, but the limit is very weak. If there are couplings to at least two different flavors, there are strong limits from LFV, especially for couplings to both electrons and muons. In this case the limits from LFV processes such as $\mu\to e \gamma$ and $\mu\to 3e$ are much stronger than DD. In the absence of one of these couplings DD limits are stronger above a certain DM mass given by the experimental energy thresholds of the DD experiments. 
In some cases the same particles entering in the DD loop may
naturally violate lepton number (specially if the DM couples to the
left-handed lepton doublets) and give rise to radiative neutrino mass models
such as the scotogenic model with Majorana DM~\cite{Ma:2006km} or the generalized scotogenic model with Dirac DM~\cite{Hagedorn:2018spx}.

(ii) If the dark fermion is a right-handed neutrino, it may be a Majorana
fermion and an active Majorana neutrino mass term is generated via the seesaw
mechanism~\cite{Minkowski:1977sc}. As the particles in the loop are neutral, DD
is generated via $Z$ and Higgs penguin diagrams~\cite{Gonzalez-Macias:2016vxy},
which are very suppressed. Although the DM may be assigned lepton flavor and
lepton number, there are no strong limits from LFV or lepton number violation
beyond those already present in seesaw scenarios. This scenario is normally
referred to as the neutrino-portal to DM~\cite{Gonzalez-Macias:2016vxy,
Escudero:2016ksa,Escudero:2016tzx}.

(iii) If the scalar is the SM Higgs, there is mixing
between the DM and the neutral component of the fermion in the loop,
which generates tree-level contributions mediated by the $Z$ boson and
the Higgs. The $Z$-mediated tree-level DD is expected to dominate with
respect to the dipole moment contributions arising at loop level. In
fact, elastic $Z$-mediated contributions are already ruled-out by DD
experiments.

While the correct relic abundance is easily achieved in models with DM couplings to SM leptons (or not too heavy right-handed neutrinos), it requires further model-building in the case of DM couplings to vector-like fermions. We have also found that the invisible loop-induced $Z$ and Higgs boson decays may sometimes impose restrictions in the case of light DM. 

In this work we studied the prototypical case of fermion singlet DM with the simplest dark sector, where the loop suppression still allows reasonably large DM interactions. Hopefully a positive DD signal in the next years will serve as a motivation and guidance to continue exploring the WIMP DD theory space and its interplay with other beyond the Standard Model probes.

\section*{Acknowledgments} 
We thank Yue-Ling Sming Tsai to provide a preliminary version of \texttt{LikeDM}~\cite{Liu:2017kmx} and for answering many questions. 
EM is grateful to Viviana Niro, Paolo Panci and Francesco Sannino for useful discussions.  JH-G acknowledges Fady Bishara for providing an earlier version of \texttt{DirectDM}~\cite{Bishara:2017nnn}. MS thanks Yi Cai for numerous useful discussions. We thank Michael Gustafsson for pointing out an error in Fig.~\ref{fig:NRVL}.
This work has been supported in part by the Australian Research Council. JH-G acknowledges the support from the Australian Research Council through the ARC Centre of
Excellence for Particle Physics at the Terascale (CoEPP) (CE110001104). All Feynman diagrams were generated using the Ti\textit{k}Z-Feynman
package for \LaTeX~\cite{Ellis:2016jkw}.

\appendix

\section{Larger dark matter groups} \label{ap:darkmattergroup}

In the main part of the text we restricted ourselves to a global
U(1) symmetry for a Dirac DM and to a discrete $Z_2$
symmetry for a Majorana DM. Our results can be easily
generalized if the DM forms a larger non-trivial representation of the dark symmetry group and there are multiple degenerate components of the DM multiplet. As the dark symmetry commutes with the SM gauge group it simply leads to an overall factor of
\begin{equation}
	\sum_\gamma	(C^{\gamma\dagger} C^\gamma)_{\alpha^\prime \alpha} \equiv \sum_{\beta,\gamma} C^{\gamma *}_{\beta \alpha'} C^\gamma_{\beta\alpha}
\end{equation}
to the Wilson coefficients of a DM particle-nucleus scattering, $\psi_\alpha N \to
\psi_{\alpha^\prime} N$,  where the Clebsch-Gordan coefficients
$C_{\beta\alpha}^\gamma$ are defined such that the scalar and the
two fermions are invariant under the dark sector symmetry:
\begin{equation}
	 C_{\beta\alpha}^\gamma \bar F_\beta S_\gamma \left(y_1 P_L+y_2 P_R\right) \psi_\alpha\;.
 \end{equation}
Thus for a general DM candidate with $N$ components $\psi_\alpha$ the DD cross section is obtained by summing over the final states and averaging over the initial state and thus
\begin{equation}
	\sigma \to \frac{\sigma}{N} \sum_{\gamma,\delta} \mathrm{Tr}\left(C^{\gamma\dagger}C^\gamma C^{\delta\dagger}C^\delta\right)\;.
\end{equation}
Note that a larger dark sector symmetry may lead to multiple DM candidates, which requires to go beyond the discussed scenario, see for instance Ref.~\cite{Herrero-Garcia:2017vrl}.

\section{Direct detection differential cross section and event rate}  \label{ap:DD}
The differential cross section for fermionic DM may be written in terms of NR operators at the nucleon level~\cite{Bishara:2017pfq}
\begin{align}\label{sigma}
	\frac{d\sigma}{dE_R}  = \frac{m_A}{2\pi v^2} \frac{4\pi}{2J_A+1} \sum_{\tau,\tau^\prime=\{0,1\}} \Big[ 
			     &R_M^{\tau\tau^\prime} W_M^{\tau\tau^\prime}(|\vec q|) 
		+ R_{\Sigma^{\prime\prime}}^{\tau\tau^\prime} W_{\Sigma^{\prime\prime}}^{\tau\tau^\prime} (|\vec q|) 
		+ R_{\Sigma^{\prime}}^{\tau\tau^\prime} W_{\Sigma^{\prime}}^{\tau\tau^\prime} (|\vec q|)
		\\\nonumber &
		+\frac{|\vec q|^2}{m_N^2} \left(
		 R_{\Delta}^{\tau\tau^\prime} W_{\Delta}^{\tau\tau^\prime} (|\vec q|)
		 + R_{\Delta\Sigma^\prime}^{\tau\tau^\prime} W_{\Delta\Sigma^\prime}^{\tau\tau^\prime} (|\vec q|)
		\right)
	\Big]
\end{align}
with the nucleus mass $m_A$ and spin $J_A$. 
The coefficients $R_X$ are given in terms of the NR Wilson
coefficients $c_i^{0,1} = (c_i^p \pm c_i^n)/2$  and $W_X$ denote the nuclear response
functions. The explicit forms of 
$R_X$ and 
$W_X$ are given in Ref.~\cite{Anand:2013yka}. For $|\vec
q|\to 0$, the long wavelength limit, $W_M(0)\propto A^2$ counts the number of
nucleons in the nucleus, $W_{\Sigma^{\prime\prime}}$ and $W_{\Sigma^\prime}$
measure the nucleon spin content of the nucleus, $W_\Delta$ measures the
nucleon angular momentum and $W_{\Delta\Sigma^\prime}$ the interference. 

In the literature it is also common to show the differential cross section as the sum of different dipole and charge contributions. Neglecting the $Z$ contributions to SD interactions, which are suppressed with respect to the long-range interactions, and taking $d_\psi=0$, the differential cross section can be written as~\cite{Banks:2010eh}:
\begin{align}
\label{eq:diffxs}
\frac{\text{d}\sigma}{\text{d}E_R} = \, &\frac{\alpha^2_{\text{em}}}{4\pi} \mu_\psi^2 Z^2 \left( \frac{1}{E_R} -\frac{m_A}{2 \mu_{\psi\text{A}}^2 v^2} \right) F^2_{\text{SI}} \left(E_R\right) \\\nonumber
 &+ \alpha^2_{\text{em}} \frac{\mu_A^2 \mu^2_\psi m_A}{4 \pi^2 v^2} \,\frac{J_A+1}{3J_A}\, F^2_{\text{SD}} \left(E_R\right) \\\nonumber
 &+ \frac{m_A}{2 \pi v^2} A^2_{\rm eff} F^2_{\text{SI}} \left(E_R\right) \,,
\end{align}
where $\mu_{\psi A}=m_\psi m_A/(m_\psi + m_A)$ is the DM--nucleus reduced mass and $A_{\rm eff} $ encodes the DM-nucleus couplings (see e.g.~Ref.~\cite{Ibarra:2015fqa}):
\begin{equation}
A_{\rm eff} = Z \left(c_{\text{SI}}^{{\rm p},Z}+
c_{\text{SI}}^{{\rm p},\gamma} +
c_{\text{SI}}^{{\rm p},H} 
- \frac{\alpha_{\text{em}} \mu_\psi}{2 \pi m_\psi} \right) 
+ (A-Z) \left(
c_{\text{SI}}^{{\rm n},Z}
+c_{\text{SI}}^{{\rm n},H}
\right)\;.
\end{equation}
The first line in Eq.~\eqref{eq:diffxs} corresponds to the dipole-charge (D-C), the second line to the dipole-dipole (D-D) and the third line to the charge-charge (C-C) interaction. $F_{\text{SI}} \left( E_R \right)$ and $F_{\text{SD}} \left(E_R\right)$ are the nuclear form factors.  $c_{\text{SI}}^{{\rm N}}$ with $N=n,p$ are the relativistic Wilson coefficients at the nucleon level for the operators
\begin{align}
	\mathcal{O}_{\rm SI}^{\rm N,V} &= \bar\psi\gamma_\mu \psi \bar N \gamma^\mu N\,,&
	\mathcal{O}_{\rm SI}^{\rm N,H} &= \bar\psi \psi \bar N N\;.
\end{align}
The vector operator $\mathcal{O}_{\rm SI}^{\rm N,V}$ is induced by both interactions with a photon and a $Z$ boson.

Once the differential cross section is computed via Eq.~\eqref{sigma}, the differential event rate per unit detector mass (for a detector with just one type of nucleus $A$) is given by:
\begin{equation}
\frac{\mathrm{d}R}{\mathrm{d}E_R} = \frac{\rho_\psi}{m_\psi m_A} \int_{v_{\rm min}(E_{\rm R})}  \frac{\mathrm{d}\sigma}{\mathrm{d}E_R} v f_{\rm det}(\vec v)\,\mathrm{d}^3 v\,,
\label{eq:difrate}
\end{equation}
where $\rho_\psi$ is the local WIMP density,  $f_{\rm det}(\vec v)$ is the WIMP velocity distribution in the detector rest frame and $v_{\rm min}$ is the minimum WIMP velocity required to produce a recoil with energy $E_R$
\begin{equation}
v_{\rm min}(E_{\rm R}) = \sqrt{\frac{E_R m_A}{2\mu_{\psi\rm A}}}\,.
\end{equation}
The velocity distribution in the detector rest frame is related to the velocity
distribution in the galaxy frame $f_{\rm gal}(\vec{v},t)$  by a simple Galilean
transformation, $f_{\rm det}(\vec{v}) = f_{\rm gal}(\vec{v} + \vec{v}_{E}(t))$,
where $\vec{v}_{E}(t)$ is the velocity of the Earth in the galactic frame.  In
our analysis we use \texttt{LikeDM} and 
refer to \cite{Huang:2016pxg,Liu:2017kmx} for the technical details of the different detectors and
astrophysical assumptions.

\mathversion{bold}
\section{Expressions for $Z$ and Higgs boson decays into dark matter} \label{ap:HZdecays} 
\mathversion{normal}

The relevant interactions of the DM $\psi$ with the Higgs and the $Z$ boson can be parameterized as\footnote{In the case of radiative neutrino mass models such as the scotogenic model~\cite{Ma:2006km} and its variants~\cite{Cai:2017jrq,Hagedorn:2018spx}, there are extra (lepton number conserving)  invisible Higgs boson decays into neutrinos at one loop, which are not suppressed by phase space and could therefore be larger than those into DM.}
\begin{equation}
\mathcal{L}_{H\psi}\;= \overline{\psi} \,\left(b_V \,+b_A \,\gamma_5 \right)  \psi \,h \,+\,\text{H.c.}\,, \label{LpsidH}
\end{equation}
and
\begin{equation}
\mathcal{L}_{Z\psi}\;= \overline{\psi} \,\left(c_V \, \gamma^\mu +c_A \, \gamma^\mu \gamma_5 +d_V \, p_2^\mu +d_A \, p_2^\mu \gamma_5 \right)\psi \,Z_\mu +\,\text{H.c.}\,,\label{LpsidZ}
\end{equation}
where $p^\mu_2$ is the 4-momentum of the outgoing DM $\psi$. We define $x_h\equiv m_\psi/m_h$ and $x_Z\equiv m_\psi/m_Z$. The partial Higgs decay width into the DM $\psi$ is non-zero for $m_\psi< m_h/2$ and reads:
\begin{equation}
	\Gamma_{ h \rightarrow \psi \psi} =\frac{S\,N_\psi m_h}{2 \pi} \left\{ [\mathrm{Re} (b_V)]^2 \left(1- 4\,x_h^2 \right) + [\mathrm{Im} (b_A)]^2 \right\} \left(1- 4\,x_h^2  \right)^{1/2}\,. 
\end{equation}
Similarly the partial width of the $Z$ is given by
\begin{eqnarray}
\begin{split}
	\Gamma_{Z \rightarrow \psi \psi} =&\; \frac{S\, N_\psi m_Z}{3 \pi}\, \left(1- 4\,x_Z^2  \right)^{1/2}\, \Big\{ [\mathrm{Re} (c_V)]^2 \left(1+2\,x_Z^2 \right) + [\mathrm{Re} (c_A)]^2  \left(1-4\,x_Z^2 \right)\\\
					      &\; +\frac{m_Z^2}{8} \,\left(1-4\,x_Z^2 \right)\,\left[[\mathrm{Re} (d_V)]^2 \left(1-4\,x_Z^2 \right)  +[\mathrm{Im}(d_A)]^2 -\frac{8x_Z}{m_Z} \mathrm{Re} (d_V) \mathrm{Re} (c_V) \right]\Big\} \label{invZ}
\end{split}	
\end{eqnarray}
for $m_\psi<m_Z/2$. $S$ is the symmetry factor, equal to $1/2$ for identical final states (Majorana DM), and equal to $1$ for Dirac DM. 
The coefficients relevant for the decays of the Higgs boson to Dirac DM can be expressed in terms of the Passarino-Veltman functions
\begin{align}
	b_V & = \frac{\lambda_{HS} v }{32\pi^2} \left[ m_F \left( |y_A|^2-|y_V|^2\right) C_0\left(m_\psi^2, m_h^2, m_\psi^2, m_F, mS, mS\right)\right.\\\nonumber
	    &\left.+2 m_\psi \left(|y_A|^2+|y_V|^2\right) C_{1}\left(m_\psi^2,m_h^2,m_\psi^2,m_F,m_S,m_S\right) \right]\,,\\
		\qquad b_A & = \frac{i\,\lambda_{HS} v}{16\pi^2} m_F\, \mathrm{Im}\left[y_Vy_A^*\right] C_0(m_\psi^2,m_h^2,m_\psi^2,m_F,m_S,m_S)\;. \label{invHc}
\end{align}
The mass insertions, $m_\psi$ and/or $m_F$ are needed in order to flip chirality.
We do not report the expressions for the decays of the $Z$ boson, as they are very long and not illustrative.

For Majorana DM $c_V=d_V=d_A=0$ and the remaining non-zero Wilson coefficients are a factor of two larger,
	$
	\left.c_A\right|_{\rm Majorana} 
		=2 \left. c_A\right|_{\rm Dirac}
			$, $
	\left.b_V\right|_{\rm Majorana} 
		=2 \left. b_V\right|_{\rm Dirac} 
			$, and $
	\left.b_A\right|_{\rm Majorana} 
		=2 \left. b_A\right|_{\rm Dirac} 
			$
due to the presence of crossed diagrams. This is analogous to
direct detection: $c^q_{\rm SS}$ and $c^q_{\rm AA}$ for Majorana
DM are a factor $2$ larger than for Dirac DM (see
Sec.~\ref{sec:ZHMaj}).

\section{Lepton flavor violation and anomalous dipole moments} \label{ap:LFV}
If the DM couples to SM leptons there may be LFV processes and anomalous electric and magnetic dipole moments. We provide the relevant expressions for DM coupling to either the left-handed SM doublets or the right-handed SM singlets. The results are identical for Dirac or Majorana DM.

\subsection{Left-handed lepton doublet}

The relevant interaction term for LFV processes is with the charged scalars:
\begin{align}
\mathcal{L}_{L_L} &= -\,  y_2 \, \overline{L_{\rm L}} \, S\,\psi_{\rm R}\,+\,\text{H.c.}
= \,  y_2 \, \overline{e_{\rm L}} \, S^-\,\psi_{\rm R}\,+\,\text{H.c.} +\dots\,.
 \label{LL}
\end{align}
The most general amplitude for the electromagnetic charged lepton flavor transition $\ell_\alpha(p)\to \ell_\beta(k)\,\gamma^*(q)$ can then be parameterized as 
 \cite{Hisano:1995cp}
\begin{equation}
	\mathcal{A}_\gamma \;=\; e\,\epsilon_\rho^*(q)\,\overline{u}(k)\Big[ q^2\,\gamma^\rho\left( A_1^L\,P_L\,+\,A_1^R\,P_R \right)\,+\,
						 m_\beta\,i\,\sigma^{\rho\sigma} \left(A_2^L\,P_L\,+\,A_2^R\,P_R\right) q_\sigma  \Big] u(p)\,,\label{ampgamma}
\end{equation}
where $e>0$ is the proton electric charge, $p$ ($k$) is the momentum of the initial (final) charged lepton $\ell_\alpha$ ($\ell_\beta$), and $q = p-k$ is the momentum of the photon.  As is well known, the charged lepton radiative decays are mediated  by the electromagnetic dipole transitions in Eq.~\eqref{ampgamma} and the corresponding branching ratio (Br) for $\ell_\alpha\to \ell_\beta\,\gamma$ is given by
\begin{equation}
	\text{Br}(\ell_\alpha\to \ell_\beta\,\gamma) \;=\; \frac{48\,\pi^3\,\alpha_{\rm em}}{G_F^2}\,\Big[\left| A_2^L\right|^2\,+\,\left| A_2^R\right|^2  \Big] 
	\times \text{Br}\left(\ell_\alpha\to\ell_\beta\,\nu_\alpha\,\overline{\nu_\beta}\right)\,.\label{BRmueg2}
\end{equation}
where
\begin{equation}\label{gammaform1}
        A_2^L \;= 0\,, \qquad \qquad
        A_2^R \;=\;   -\frac{1}{32\,\pi^2}\,\frac{ y_2^{\beta} \, y_2^{\alpha*}}{m_{S^\pm}^2}\,f\left(\frac{m_\psi^2}{m_{S^\pm}^2}\right)\,,
\end{equation}
with
\begin{equation}\label{f-loop}
	f(x)  \; = \; \frac{1-6x+3x^2+2x^3-6x^2\log(x)}{6(1-x)^4}\,. 
\end{equation}
For trilepton decays we consider only the contributions from the photon penguin
and from box-type diagrams, as the $Z$ penguin is suppressed by charged lepton
masses. Box diagrams may be the dominant contribution in absence of the
contributions from photon and $Z$ penguins. The amplitude from the box diagrams
is given by
\begin{equation} \label{eq:box}
	\mathcal{A}_{\text{BOX}} \;=\; e^2 B\, \overline{u}(k_1)\,\gamma^\alpha\,P_L\,u(p)\,
	\overline{u}(k_3)\,\gamma_\alpha\,P_L\,v(k_2)\,.
\end{equation}
For same-flavor leptons in the final state the branching ratio of $\ell_\alpha \to \ell_\beta \, \overline{\ell_\beta}\, \ell_\beta$  reads:
\begin{equation}\label{ell3ellBR}
\begin{split}
\text{Br}(\ell_\alpha \to \ell_\beta \, \overline{\ell_\beta} \, \ell_\beta)\;=\;&\frac{6\pi^2\alpha_{\text{em}}^2}{G_F^2} \Bigg[ 
\left|A_1^L\right|^2+\left|A_2^R\right|^2\left(\frac{16}{3}\ln\frac{m_\alpha}{m_\beta}-\frac{22}{3}\right) \\
&\;+\frac16\left|B\right|^2 -4\,\text{Re} \left(A_1^{L*} A_2^{R}-\frac16\left(A_1^L -2A_2^R \right)B^{*}\right)\Bigg]\\
&\,\times \text{Br}\left(\ell_\alpha\to\ell_\beta\,\nu_\alpha\,\overline{\nu_\beta}\right)\,.
\end{split}
\end{equation}
For $\ell_\alpha^- \to \ell_\beta^- \ell_\gamma^- \ell_\gamma^+$ with $\beta \ne \gamma$ the branching ratio reads:
\begin{equation}\label{ell3ellBRb}
\begin{split}
\text{Br}(\ell_\alpha \to \ell_\beta \, \overline{\ell_\gamma} \, \ell_\gamma)\;=\;&\frac{6\pi^2\alpha_{\text{em}}^2}{G_F^2} \Bigg[ 
\frac{2}{3}\left|A_1^L\right|^2+\left|A_2^R\right|^2\left(\frac{16}{3}\ln\frac{m_\alpha}{m_\gamma}-8\right) \\
&\;+\frac{1}{12}\left|B\right|^2 -\frac{8}{3}\,\text{Re} \left(A_1^L A_2^{R*}-\frac18\left(A_1^L -2A_2^R \right)B^{*}\right)\Bigg]\\
&\,\times \text{Br}\left(\ell_\alpha\to\ell_\beta\,\nu_\alpha\,\overline{\nu_\beta}\right)\,.
\end{split}
\end{equation}
For $\ell_\alpha^- \to \ell_\beta^+ \ell_\gamma^- \ell_\gamma^-$ we get
\begin{equation}\label{ell3ellBRc}
\begin{split}
\text{Br}(\ell_\alpha \to \overline{\ell_\beta} \, \ell_\gamma \, \ell_\gamma)\;=\;&\frac{\pi^2\alpha_{\text{em}}^2}{G_F^2}
\left|B\right|^2 \times \text{Br}\left(\ell_\alpha\to\ell_\beta\,\nu_\alpha\,\overline{\nu_\beta}\right)\,,
\end{split}
\end{equation}
because there are only contributions from box diagrams.
The coefficients $A_2^{L,\,R} $ are given in Eq.~\eqref{gammaform1} and 
\begin{equation}\label{gammaform2}
	A_1^L \;= \; -\frac{1}{48\,\pi^2}\,\frac{ y_2^{\beta} \, y_2^{\alpha*}}{m_{S^\pm}^2}\,g\left(\frac{m_\psi^2}{m_{S^\pm}^2}\right)\,\,, \qquad \qquad A_1^R \;= 0\; \,,
\end{equation}
with
\begin{equation}\label{g-loop}
	g(x) \; = \; \frac{2-9x+18x^2-11x^3+6x^3 \log(x)}{12(1-x)^4} \,. 
\end{equation}
The contribution from box diagrams $B$ for $\ell_\alpha^- \to \ell_\beta^- \ell_\gamma^- \ell_\gamma^+$ reads
\begin{equation} \label{eq:B1}
	e^2 B\;=\; \frac{1}{16\,\pi^2}\Bigg[\frac{y_2^{\alpha*} y_2^{\beta} y_2^{\gamma}y_2^{\gamma*}}{m_{S^\pm}^2}\,h\left(\frac{m_\psi^2}{m_{S^\pm}^2}\right) \Bigg]\,,
\end{equation}
and for $\ell_\alpha^- \to \ell_\gamma^- \ell_\gamma^- \ell_\beta^+$ it is given by
\begin{equation} \label{eq:B2}
	e^2 B\;=\;  \frac{1}{16\,\pi^2}\Bigg[\frac{y_2^{\alpha*} y_2^{\beta*} (y_2^{\gamma})^2}{m_{S^\pm}^2}\,h\left(\frac{m_\psi^2}{m_{S^\pm}^2}\right) \Bigg]
\end{equation}
with
\begin{equation}
	h(x) \;=\;  \frac{1-x^2+2x\ln x}{2(x-1)^3}\,.
\end{equation}
All the external momenta and masses have been neglected. Of course for $\ell_\alpha \to \ell_\beta \, \overline{\ell_\beta}\, \ell_\beta$ both Eq.~\eqref{eq:B1} and Eq.~\eqref{eq:B2} agree with $\gamma=\beta$.

For $\mu-e$ conversion in nuclei we only consider coherent scattering via photon contributions, but include both short- and long-range contributions \cite{Kitano:2002mt}:\footnote{We neglect the $Z$ boson contribution which is proportional to the square of the charged lepton masses and thus negligible compared to the photon penguin diagram.}
\begin{eqnarray}
    {\cal L}_{\rm int} &=&
- \frac e2
    \left(
    m_\mu A_2^L\, \overline{\ell_e}\, \sigma^{\mu \nu} P_L\, \ell_\mu F_{\mu \nu}
    + m_\mu A_2^R\, \overline{\ell_e}\, \sigma^{\mu \nu} P_R\, \ell_\mu F_{\mu \nu}
    + {\rm h.c.}
    \right) \nonumber \\
    &&
    - 
 \sum_{q = u,d,s}
    \left[ {\rule[-3mm]{0mm}{10mm}\ } \right.
    \left(
    g_{LV(q)}^\gamma\, \overline{\ell_e} \gamma^{\alpha} P_L \ell_\mu
   + g_{RV(q)}\, \overline{\ell_e} \gamma^{\alpha} P_R \ell_\mu
   \right) \overline{q} \gamma_{\alpha} q
	+ {\rm h.c.}
    \left. {\rule[-3mm]{0mm}{10mm}\ } \right].\;\;\;\;\;\;\;
    \label{eq:mue}
\end{eqnarray}
The $\mu-e$ conversion rate is%
\begin{align}\label{eq:mu2eOmega}
	\omega_{\rm conv} \;=\;&
	4\,\left| \frac{e}{8} A_2^{R} D+ \tilde{g}_{LV}^{(p)} V^{(p)} + \tilde{g}_{LV}^{(n)} V^{(n)}\right|^2  \ ,
\end{align}
where the effective vector couplings $\tilde{g}_{L/RV}^{(p,n)}$ for the proton and the neutron are 
\begin{align}
\tilde g_{LV}^{(p)}&\approx 2\, g_{LV(u)}^\gamma + g_{LV(d)}^\gamma = e^2 A_1^L\ ,  &  \tilde g_{LV}^{(n)}&\approx g_{LV(u)}^\gamma + 2\, g_{LV(d)}^\gamma =0\,,
\end{align}
with 
\begin{equation}
\begin{split}
	g_{LV(q)}^\gamma \; = \; & e^2\, Q_q\, A_1^L\,.
	\end{split}
\end{equation}
The coefficients $A_1^{L,\,R} $ are given in Eq.~\eqref{gammaform2}, and $Q_q$ is the quark electric charge of the quark $q$ in units of $e>0$. The numerical values of the overlap integrals $D$ and $V^{(p,n)}$ and the total capture rate for each nucleus are reported in Tab.~\ref{tab:overlap} for three different nuclei. As we only consider the photon contribution and thus only couplings to the electric charge of the quarks, there is no effective coupling to neutrons.
\begin{table}[tbp]
\centering\setlength{\extrarowheight}{3pt}
\begin{tabular}{|c|ccc|c|}
\hline
 &  $V^{(p)}$ & $V^{(n)}$ & $D$ & $\omega_{\rm capt}(10^6 s^{-1})$\\
 \hline\hline
 $^{197}_{79}\text{Au}$ & 0.0859 & 0.108  & 0.167 & 13.07 \\
 $^{48}_{22}\text{Ti}$  & 0.0399 & 0.0495 & 0.0870  & 2.59\\
$^{27}_{13}\text{Al}$   & 0.0159 & 0.0169 &  0.0357 & 0.7054 \\
\hline
\end{tabular}
\vspace{1ex}

\begin{minipage}{14cm}
\caption{The overlap integrals in the units of $m_{\mu}^{5/2}$ and the total capture rates for different nuclei \cite{Kitano:2002mt}. 
	The total capture rates are taken from Tab.~8 in~\cite{Kitano:2002mt}. 
The overlap integrals of $^{197}_{79}\text{Au}$ as well as $^{27}_{13}\text{Al}$ are taken from Tab. 2 and 
$^{48}_{22}\text{Ti}$ are taken from Tab.~4 of Ref.~\cite{Kitano:2002mt}.
 \label{tab:overlap}}
\end{minipage}
\end{table}

Even if lepton flavor is conserved there are processes that can bound the DM interactions with the leptons. Electric dipole moments for the leptons occur in these simplified models only at the two-loop level. However leptonic magnetic dipole moments occur at one-loop order via photon penguin diagrams, similarly to $\mu \rightarrow e \gamma$ transitions. They receive two independent contributions from the charged scalars running in the loop, which are given by~\cite{Raidal:2008jk}:
\begin{equation}
	\Delta a_\ell \equiv \frac{g_\ell-2}{2} \;=\; m^2_\ell\, {\rm Re} [A_2^R]_\ell\,.
	\label{gminus2}
\end{equation}
$A_2^R$ is the diagonal part ($\alpha=\beta \equiv \ell$) of the coefficient given in Eq.~\eqref{gammaform1} and the loop function is defined in Eq.~\eqref{f-loop}. Our expression agrees with Ref.~\cite{Kopp:2009et}. In the case of the muon magnetic dipole moment, the discrepancy with the SM has the opposite sign and hence the model cannot explain it. However this can be used to (very weakly) bound the model. Electron and tau AMMs do not lead to any relevant constraints.

\mathversion{bold}
\subsection{Right-handed charged lepton}
\mathversion{normal}

The relevant interaction term for LFV processes is with the charged scalars:
\begin{align}
\mathcal{L}_{L_L} &= -\, y_1 \, \overline{e_{\rm R}} \, S^-\,\psi_{\rm L}\,+\,\text{H.c.}\,.
 \label{LeR}
\end{align}
All the expressions are the same as for the left-handed lepton doublets after substituting the right-handed superscript by the left-handed one, i.e., $A_{1,\,2}^R \leftrightarrow A_{1,\,2}^L$, $g^\gamma_{LV} \leftrightarrow g_{RV}^\gamma$ and the Yukawa couplings $y_1 \leftrightarrow y_2$.

\section{Computation of the relic abundance} \label{ap:relic_details}
In this appendix we review the computation of the relic abundance, see for instance Refs.~\cite{Kolb:1990vq,Griest:1990kh}. We use the instantaneous freeze-out approximation which is sufficient for our purposes. The final DM abundance is determined by
\begin{equation} \label{eq:Omega}
	\Omega_\psi = \frac{n+1}{\lambda} x_f^{n+1} \frac{m_\psi s_0}{\rho_{\rm cr}}\,,
\end{equation}
with $\lambda=\left[x  s \braket{\sigma v}/H\right]_{x=1}$ and $n=0\, (1)$ for s-wave (p-wave) DM annihilation. The entropy density is denoted by $s$, with today's value given in terms of the CMB temperature $T_{\gamma,0}=2.73$ K as $s_0 = 2\pi^2/45\, (43/11)\,T_{\gamma,0}^3$, where we have used $N_{\rm eff}=3$.

Equating the interaction rate $\Gamma_{\rm ann}$ for the process $\psi\bar\psi \leftrightarrow \ell_\alpha \bar\ell_\beta$ with the Hubble rate, $H (T_f)=\Gamma_{\rm ann} (T_f)$ we obtain a condition for the freeze-out temperature
\begin{equation} \label{eq:FO}
	\sqrt{\frac{\pi^2}{90\, m_{\rm P}^2} g_*} = \frac{g_\psi m_\psi \braket{\sigma v}}{(2\pi)^{3/2}} x_f^{1/2} e^{-x_f}\,,
\end{equation}
where $m_{\rm P}$ is the Planck mass, $g_*$ is the number of relativistic degrees of freedom at freeze-out ($g_*=106.75$ in the SM), and $g_\psi$ is the DM number of degrees of freedom, which is equal to $2\, (4)$ for Majorana (Dirac) DM.

The annihilation cross section may implicitly depend on the freeze-out temperature, and it is useful to factorize out this dependence. Then Eq.~\eqref{eq:FO} can be written in terms of $\lambda$ as
\begin{equation} \label{eq:FO2}
	\frac{4}{3} \frac{\pi^2}{30} \frac{g_*^s}{g_\psi} \frac{(2\pi)^{3/2}}{\lambda} = x_f^{\frac12-n} e^{-x_f}\,.
\end{equation}
Solving for $\lambda$ in Eq.~\eqref{eq:FO2}, plugging it in Eq.~\eqref{eq:Omega}, and imposing that relic abundance matches the observed value $\Omega_\psi h^2=0.12$~\cite{Ade:2015xua}, one can numerically obtain the value of $x_f$. We get values of $23 \lesssim x_f \lesssim 30$ for $ 10\,\text{GeV}\lesssim m_\psi \lesssim 10^4\,\text{GeV}$, which turn out to be identical for Dirac and for Majorana DM. We also note that $x_f$ increases roughly logarithmically with the DM mass $m_\psi$. 

For a given $(m_\psi, x_f)$ pair Eq.~\eqref{eq:FO2} allows one to compute the
annihilation cross section averaged over velocity, $\braket{\sigma v}$, which
depends exponentially on $x_f$. For the range of DM masses given above the dependence on the DM mass is very mild. We
obtain that the required thermally averaged annihilation cross sections to
reproduce the observed DM abundance are in the range $1.8 \lesssim
10^{26}\left.\braket{\sigma v}\right|_\text{D}  ({\rm cm^3\, s^{-1}})\lesssim
	2.4$ and $4 \lesssim 10^{24}\left.\braket{\sigma v}\right|_\text{M}
		({\rm cm^3\, s^{-1}})\lesssim 9$ for Dirac and Majorana DM, respectively.

\section{Matching onto non-relativistic operators}\label{ap:NRmatching}

We use \texttt{DirectDM}~\cite{Bishara:2017nnn} which follows the normalization of the NR operators in Ref.~\cite{Anand:2013yka} to match our one-loop calculation of DM scattering off quarks onto the NR operators using 3 flavor QCD without running, i.e. the matching occurs at $\mu=2$ GeV. This is justified as the relevant relativistic operators are renormalization group invariant under one-loop QCD corrections. There are no additional significant contributions, because the particles in the loop are color singlets. There can be sizable renormalization group corrections, if there are colored particles in the loop, see e.g.~the discussion of bino DM in the minimal supersymmetric SM in Ref.~\cite{Berlin:2015njh}. 

Note that the coefficients $c^q_i$  depend on the 3-momentum transfer $|\vec q|=\sqrt{2\,m_A E_R}$ with the target nucleus mass $m_A$ and the recoil energy $E_R$. In the numerical examples in the figures we use $E_R=8.59$ keV for ${}^{132}_{54}\mathrm{Xe}$ which results in $|\vec q|^2=2.11\times 10^{-3}\,\mathrm{GeV}^2$. The exact numerical expressions used in the code are given below. All quantities are defined in units of GeV.
All NR Wilson coefficients have dimension $\mathrm{GeV}^{-2}$.
Higgs penguins with heavy SM quarks $Q$ are described by the Wilson coefficient of the gluon operator 
	\begin{equation}
		c_g=-\sum_{Q=t,b,c} c^Q_{\text{SS}}\;.
	\end{equation}

	\subsection{Dirac dark matter}
NR Wilson coefficients for protons
\begin{align}
	c_1^p&=0.032 c_{\text{SS}}^d+ c_{\text{VV}}^d-0.0628148 c_g+0.0413 c_{\text{SS}}^s+0.017 c_{\text{SS}}^u+2 c_{\text{VV}}^u-\frac{0.00119243 \mu _{\psi }}{m_{\psi }}\\
	c_4^p&=1.504 c_{\text{AA}}^d+0.124 c_{\text{AA}}^s-3.588 c_{\text{AA}}^u-0.0141733 \mu _{\psi }\\
	c_5^p&=\frac{0.00447838 \mu _{\psi }}{|\vec q|^2}\\
	c_6^p&=-\frac{2.24324 c_{\text{AA}}^d}{|\vec q|^2+0.0182187}+\frac{0.342636 c_{\text{AA}}^d}{|\vec q|^2+0.300153}-\frac{0.685272 c_{\text{AA}}^s}{|\vec q|^2+0.300153}
	\\\nonumber&
	+\frac{2.24324 c_{\text{AA}}^u}{|\vec q|^2+0.0182187}+\frac{0.342636 c_{\text{AA}}^u}{|\vec q|^2+0.300153}+\frac{0.0124947 \mu _{\psi }}{|\vec q|^2}\\
	c_{11}^p&=\frac{0.00447838 d_{\psi }}{|\vec q|^2}
\end{align}
and neutrons
\begin{align}
	c_1^n&=0.036 c_{\text{SS}}^d+2 c_{\text{VV}}^d-0.0628148 c_g+0.0413 c_{\text{SS}}^s+0.015 c_{\text{SS}}^u+ c_{\text{VV}}^u\\
	c_4^n&=-3.588 c_{\text{AA}}^d+0.124 c_{\text{AA}}^s+1.504 c_{\text{AA}}^u+0.00970284 \mu _{\psi }\\
	c_6^n&=\frac{2.24324 c_{\text{AA}}^d}{|\vec q|^2+0.0182187}+\frac{0.342636 c_{\text{AA}}^d}{|\vec q|^2+0.300153}-\frac{0.685272 c_{\text{AA}}^s}{|\vec q|^2+0.300153}
	\\\nonumber&
	-\frac{2.24324 c_{\text{AA}}^u}{|\vec q|^2+0.0182187}+\frac{0.342636 c_{\text{AA}}^u}{|\vec q|^2+0.300153}-\frac{0.00855371 \mu _{\psi }}{|\vec q|^2}
\end{align}
\subsection{Majorana dark matter}
NR Wilson coefficients for protons
\begin{align}
	c_1^p&=0.064 c_{\text{SS}}^d-0.12563 c_g+0.0826 c_{\text{SS}}^s+0.034 c_{\text{SS}}^u\\
	c_4^p&=3.008 c_{\text{AA}}^d+0.248 c_{\text{AA}}^s-7.176 c_{\text{AA}}^u\\
	c_6^p&=-\frac{4.48648 c_{\text{AA}}^d}{|\vec q|^2+0.0182187}+\frac{0.685272 c_{\text{AA}}^d}{|\vec q|^2+0.300153}-\frac{1.37054 c_{\text{AA}}^s}{|\vec q|^2+0.300153}
	+\frac{4.48648 c_{\text{AA}}^u}{|\vec q|^2+0.0182187}+\frac{0.685272 c_{\text{AA}}^u}{|\vec q|^2+0.300153}\\
	c_8^p&=4 \left(c_{\text{AV}}^d+2 c_{\text{AV}}^u\right)\\
	c_9^p&=-4.12 c_{\text{AV}}^d+0.876 c_{\text{AV}}^s+14.72 c_{\text{AV}}^u
\end{align}
and neutrons
\begin{align}
	c_1^n&=0.072 c_{\text{SS}}^d-0.12563 c_g+0.0826 c_{\text{SS}}^s+0.03 c_{\text{SS}}^u\\
	c_4^n&=-7.176 c_{\text{AA}}^d+0.248 c_{\text{AA}}^s+3.008 c_{\text{AA}}^u\\
	c_6^n&=\frac{4.48648 c_{\text{AA}}^d}{|\vec q|^2+0.0182187}+\frac{0.685272 c_{\text{AA}}^d}{|\vec q|^2+0.300153}-\frac{1.37054 c_{\text{AA}}^s}{|\vec q|^2+0.300153}
	-\frac{4.48648 c_{\text{AA}}^u}{|\vec q|^2+0.0182187}+\frac{0.685272 c_{\text{AA}}^u}{|\vec q|^2+0.300153}\\
	c_8^n&=4 \left(2 c_{\text{AV}}^d+c_{\text{AV}}^u\right)\\
	c_9^n&=14.72 c_{\text{AV}}^d+0.876 c_{\text{AV}}^s-4.12 c_{\text{AV}}^u
\end{align}

\bibliographystyle{JHEP}
\small
\setlength{\bibsep}{0pt}
\bibliography{oneloopDD}

\end{document}